\documentclass[aps,prl,twocolumn,superscriptaddress,showpacs,floatfix,longbibliography]{revtex4-2}
\usepackage{amsmath,amssymb,amsfonts,float,graphics,epsfig,epstopdf,color,verbatim,tabularx,bm,multirow,appendix,hyperref}
\usepackage{lmodern}
\usepackage{color}
\usepackage{bm}

\renewcommand{\(}{\left(}
\renewcommand{\)}{\right)}

\def\bk{{\mathbf{k}}}
\def\bK{{\mathbf{K}}}

\def\br{{\mathbf{r}}}

\def\bq{{\mathbf{q}}}

\def\bG{{\mathbf{G}}}
\def\bkp{{ \mathbf{k}^{\prime} }}

\begin{document}
	\title{Dynamical properties of collective excitations in twisted bilayer Graphene }

	\author{Gaopei Pan}
	\affiliation{Beijing National Laboratory for Condensed Matter Physics and Institute of Physics, Chinese Academy of Sciences, Beijing 100190, China}
	\affiliation{School of Physical Sciences, University of Chinese Academy of Sciences, Beijing 100190, China}
	\author{Xu Zhang}
	\affiliation{Department of Physics and HKU-UCAS Joint Institute of Theoretical and Computational Physics, The University of Hong Kong, Pokfulam Road, Hong Kong SAR, China}
	\author{Heqiu Li}
	\affiliation{Department of Physics, University of Michigan, Ann Arbor, Michigan 48109, USA}
	\affiliation{Department of Physics, University of Toronto, Toronto, Ontario M5S 1A7, Canada}
	\author{Kai Sun}
	\email{sunkai@umich.edu}
	\affiliation{Department of Physics, University of Michigan, Ann Arbor, Michigan 48109, USA}
	\author{Zi Yang Meng}
	\email{zymeng@hku.hk}
	\affiliation{Department of Physics and HKU-UCAS Joint Institute of Theoretical and Computational Physics, The University of Hong Kong, Pokfulam Road, Hong Kong SAR, China}
	
\begin{abstract}
Employing the recently developed momentum-space quantum Monte Carlo scheme, we study the dynamic response of single-particle and collective excitations in realistic continuum models of twisted bilayer graphene. At charge neutrality with small flat band dispersion, this unbiased numerical method reveals single-particle spectra and collective excitations at finite temperature. Single-particle spectra indicate that repulsive interactions push the fermion spectral weight away from the Fermi energy and open up an insulating gap. The spectra of collective excitations suggest an approximate valley $SU(2)$ symmetry. At low-energy, long-lived valley waves are observed, which resemble spin waves of Heisenberg ferromagnetism. At high-energy, these sharp modes quickly become over-damped, when their energy reaches the fermion particle-hole continuum.
	\end{abstract}

\date{\today}
\maketitle
	
{\it Introduction}\,---\, To understand the rich physics in twisted bilayer graphene (TBG), as well as the mechanism that governs this novel quantum system, a crucial step is to identify the ground state and to characterize the associated low-energy excitations~\cite{Trambly2010,Trambly2012,Bistritzer_TBG,ROZHKOV20161,Santos2007,Santos2012,cao2018unconventional,cao2018correlated,chen2020tunable,kerelsky2019maximized,Tomarken2019,lu2019superconductors,xie2019spectroscopic,Shen2020DTBG,Nuckolls_2020,pierce2021unconventional,moriyama2019,Rozen2021,liu2020spectroscopy,KhalafSoftmodes2020,Zaletel2020,Khalaf2021,Chatterjee2020}. Recently, many new insights have been obtained using real-space effective model analysis and large-scale numerical simulations (e.g. quantum Monte Carlo and DMRG)~\cite{Koshino2018,Kang2018,XYXu2018,Kang2019,YDLiao2021PRX,YDLiao2021CPB,BBChen2020}, which indicate that even at integer fillings, correlation effects give rise to a very rich phase diagram with a variety of competing quantum phases. A key advantage of this approach is that these lattice models can be easily incorporated with well-established numerical techniques, but it remains a challenge to determine the effective control parameters utilized in these models from first principle. Another parallel approach utilizes continuum models with flat bands and fragile topology~\cite{Ashvin2018,HCPo2018Fragile,HCPo2019}, where Coulomb interactions and first principle material parameters can be easily incorporated. In this approach, a key theoretical challenge is to handle the strong Coulomb interactions. In certain special limit, exact solutions exist due to emergent high symmetry~\cite{bernevig2020tbg5}. For realistic material parameters away from these special cases, Hartree-Fock mean-field and DMRG calculations suggest that the ground state is likely to be an intervalley coherent (IVC) state~\cite{Bultinck2020,YiZhang2020,JKang2020RG,Bultinck2021,KhalafSoftmodes2020,Zaletel2020,Khalaf2021,Chatterjee2020}, which mixes electron states from the two opposite valleys and breaks the $U_{v}(1)$ valley charge conservation. There have been many studies about symmetry-breaking ground states of such systems~\cite{JPLiu2019Pseudo,JPLiu2021,YiZhang2020,lian2020tbg4,Kwan2021}. While finite temperature results and the collective excitation is a matter of widespread concern. To fully understand such a complex many-body system, unbiased numerical methodology, which can solve such correlated problems efficiently and accurately, is in great need.

In this Letter, we utilize the momentum-space quantum Monte Carlo (QMC) method~\cite{Ippoliti2018,ZHLiuEMUS2019,XuZhang2021,JYLee2021} to achieve this objective. The implementation of this method in continuum models of TBG has been developed recently~\cite{XuZhang2021,JYLee2021}, but dynamic response, in particular the spectral information of the collective excitations, has not yet been obtained. In this work, we employ the momentum space QMC method, accompanied by the stochastic analytic (SAC) continuation scheme~\cite{Sandvik2016,HShao2017,GYSun2018,NSMa2018,zhou2020amplitude,ZYan2021,zhou2020amplitude,hu2020evidence}, to compute the spectra of both single-particle and particle-hole excitations. We find that, at the charge neutrality point (CNP), the IVC state is the leading instability, with strong competition from the VP state. More interestingly, although the valley $SU(2)$ symmetry is broken explicitly when control parameters take realistic values (with kinetic term), dynamic response of particle-hole excitations still exhibits an approximate $SU(2)$ symmetry. At low-energy, long-lived valley waves are observed in close analogy to spin waves of a Heisenberg ferromagnet, and these modes become over-damped as their energy reaches the particle-hole continuum. These results reveal complex dynamic response in TBG and provide a foundation for the study of other intriguing  physics at and away from charge neutrality, such as the mechanism of superconductivity and its possible topological origin~\cite{Khalaf2021,Chatterjee2020,Saito2021,Rozen2021}.

\begin{figure*}
\includegraphics[width=\textwidth]{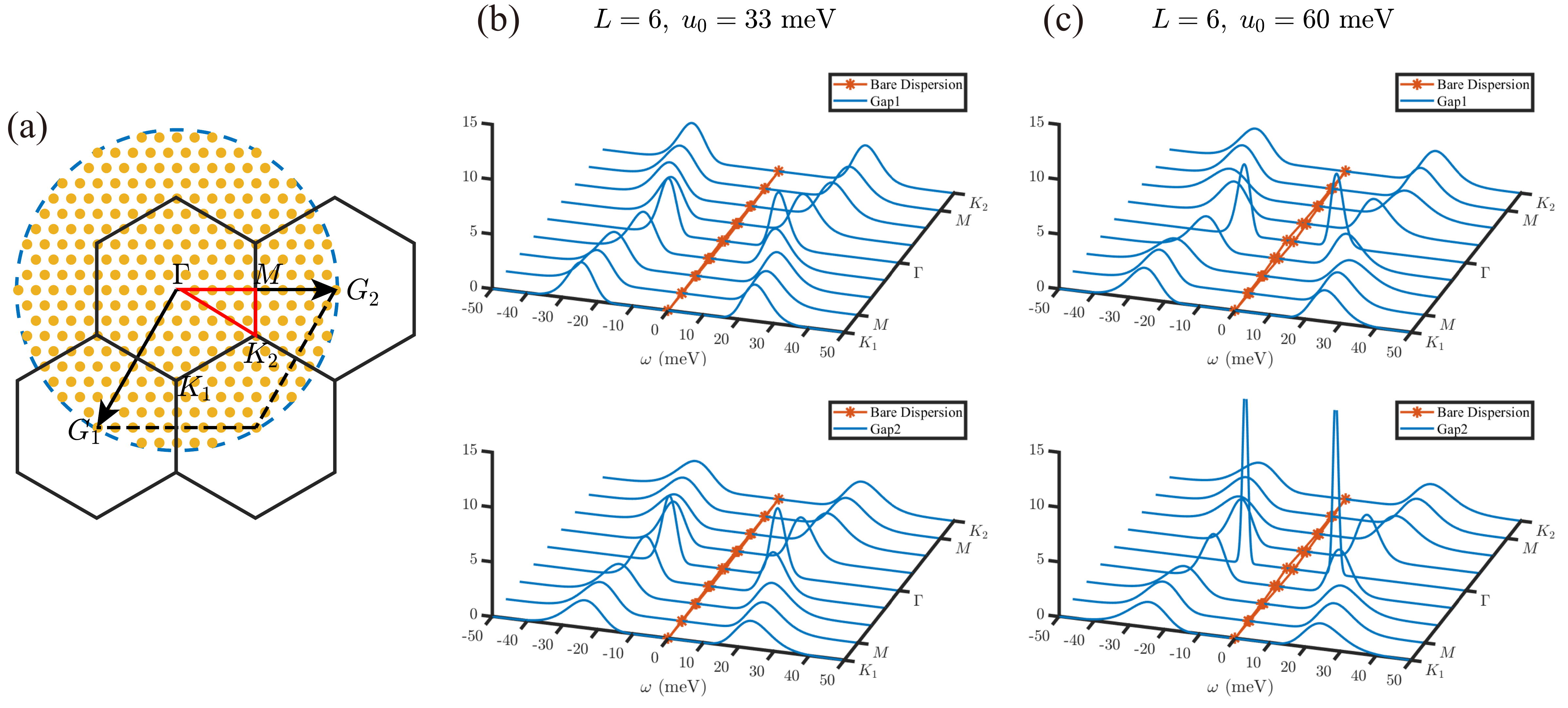}
\caption{(a) The moir\'e Brillouin zones (mBZ) at one valley. The red solid line marks the high-symmetry path $\Gamma-M-K_1(K_2)-\Gamma$. $\bG_1$ and $\bG_2$ are the reciprocal lattice vectors of the mBZ. Yellow dots mark possible momentum transfer in QMC simulations, $\bq+\bG$, and the blue dashed circle is the momentum space cut-off. Because the form factor decays exponentially with $\bG$~\cite{bernevig2020tbg5}, scatterings with momentum transfer larger than this cut-off are ignored. Here we show a $9\times 9$ mesh in the mBZ, with 300 allowed momentum transfers. In (b) and (c), blue lines are single particle spectra of $L=6,\; T =0.667$ meV, $u_0=33 $ meV and $60$ meV(realistic case \cite{YiZhang2020,bernevig2020tbg1,song2020tbg2,bernevig2020tbg3,Ashvin2019}), respectively, obtained from the momentum space QMC with analytic continuation. The red stars and lines indicate the bare dispersions of $H_0$, which is the kinetic energy in our model in Eq.(3).} 
	\label{fig:fig1}
\end{figure*}
	
{\it Model and Method}\,---\,
In this study, we utilize the continuum model of TBG flat band introduced in Refs.~\cite{Trambly2010,Trambly2012,Bistritzer_TBG,ROZHKOV20161,Santos2007,Santos2012}. In the plane wave basis, the single-particle Hamiltonian can be written as:
\begin{equation}
\begin{aligned}
H^{\tau}_{0,\mathbf{k},\mathbf{k {}^\prime}}=\delta_{\mathbf{k},\mathbf{k {}^\prime}}&\left(\begin{array}{cc} 
-\hbar v_F ({\bk}-\bK_1^{\tau}) \cdot \pmb{\sigma}^{\tau}  &  U_0  \\
U_0^\dagger  & -\hbar v_F ({\bk}-\bK_2^{\tau}) \cdot \pmb{\sigma}^{\tau}
\end{array}\right) \\ 
+&\left(\begin{array}{cc} 
0  &  U_1^{\tau} \delta_{\mathbf{k},\mathbf{k {}^\prime}-\tau \bG_1 }  \\ 
U_1^{\tau \dagger} \delta_{\mathbf{k},\mathbf{k {}^\prime}+\tau \bG_1 }  & 0
\end{array}\right)\\ 
+&\left(\begin{array}{cc} 
0  &  U_2^{\tau} \delta_{\mathbf{k},\mathbf{k {}^\prime}-\tau(\bG_1+\bG_2) }  \\ 
U_2^{\tau \dagger} \delta_{\mathbf{k},\mathbf{k {}^\prime}+\tau(\bG_1+\bG_2) }  & 0
\end{array}\right)
\end{aligned}
\label{eq:eq1}
\end{equation}
where $v_F$ is the Dirac velocity, $\tau =\pm$ is the valley index, and $\boldsymbol{\sigma}^{\tau}=(\tau \sigma_x,\sigma_y)$ defines the A,B sublattices of the monolayer graphene. And $\bK^{\tau}_{1,2}$ are the corresponding Dirac points of the bottom and top layers, which are twisted by angles $\mp\frac{\theta}{2}$ respectively. 
As shown in Fig.~\ref{fig:fig1} (a),  $\mathbf{G}_{1}=\left(-\frac{2 \pi}{\sqrt{3}L_{M}},-\frac{2 \pi}{L_{M}}\right)$ and $\mathbf{G}_{2}=\left(\frac{4 \pi}{\sqrt{3} L_{M}}, 0\right)$ are reciprocal lattice vectors of the moir\'e Brillouin zone (mBZ), with $L_{M}=a_{0} /[2 \sin (\theta / 2)]$ and $a_{0}=0.246 \mathrm{~nm}$. Interlayer tunnelings are described by $U_0=
\left(\begin{array}{cc} 
	u_0  & u_1 \\ 
	u_1 & u_0
\end{array}\right)$, $U_1^{\tau}=   \left(\begin{array}{cc} 
u_0  & u_1 e^{-\tau \frac{2\pi}{3}i} \\ 
u_1 e^{\tau \frac{2\pi}{3}i} & u_0
\end{array}\right)$ and $U_2^{\tau}= \left(\begin{array}{cc} 
u_0  & u_1 e^{\tau \frac{2\pi}{3}i} \\ 
u_1 e^{-\tau \frac{2\pi}{3}i} & u_0
\end{array}\right)$
where $u_0$ and $u_1$ are the intra- and inter-sublattice interlayer tunneling amplitudes. In this Letter, we set $\hbar v_{F} / a_{0}=2377.45$ meV, $\theta=1.08^{\circ}$ and $u_{1}= 110$ meV, which means the moir\'e bands are completely flat at the chiral limit $u_0=0$~\cite{bernevig2020tbg1,song2020tbg2,bernevig2020tbg3,Ashvin2019}.

We then project the charge-density operator at $\mathbf{q}+\mathbf{G}$ to the nearly flat bands relative to the filling of CNP:
\begin{equation}
\begin{aligned}
\delta \rho_{\mathbf{q}+\mathbf{G}}&=\sum_{\mathbf{k} \in m B Z, m_{1}, m_{2}, \tau, s} \lambda_{m_{1}, m_{2}, \tau}(\mathbf{k}, \mathbf{k}+\mathbf{q}+\mathbf{G})\\
&\qquad \left(d_{\mathbf{k}, m_{1}, \tau, s}^{\dagger} d_{\mathbf{k}+\mathbf{q}, m_{2}, \tau, s}-\frac{1}{2} \delta_{\mathbf{q}, 0} \delta_{m_{1}, m_{2}}\right)\\
&=\left(\delta \rho_{-\mathbf{q}-\mathbf{G}}\right)^{\dagger}
\end{aligned}
\end{equation}
where $d_{\boldsymbol{k}, m.\tau,s}^{\dagger}$ is the creation operator for a Bloch eigenstate, $\left|u_{\boldsymbol{k}, m,\tau,s}\right\rangle$, with $m$, $s$, $\tau$ band, spin and valley indices. The form factor is defined as $\lambda_{m_{1}, m_{2}, \tau}(\mathbf{k}, \mathbf{k}+\mathbf{q}+\mathbf{G}) \equiv\left\langle u_{\boldsymbol{k}, m_1,\tau} \mid u_{\boldsymbol{k}+\bq+\bG, m_2,\tau}\right\rangle$. As shown in Fig.~\ref{fig:fig1} (a) $\mathbf{q} \in \mathrm{mBZ}$ and $\mathbf{q}+\mathbf{G}$ represents a vector in extended mBZ, with $\bG=n_1\mathbf{G}_{1}+n_2 \mathbf{G}_{2}, \; n_1,n_2 \in \mathrm{Z}$ ~\cite{song2020tbg2,bernevig2020tbg3}.
After projecting to the flat band, the Hamiltonian reads:
	\begin{equation}
	\begin{aligned}
	H&=H_0+H_{int}\\
	H_{0}&=\sum_{m=\pm 1} \sum_{\mathbf{k} \tau s} \epsilon_{m, \tau}(\mathbf{k}) d_{\mathbf{k},m, \tau, s}^{\dagger} d_{\mathbf{k}, m, \tau, s} \\
	H_{i n t}&=\frac{1}{2 \Omega} \sum_{\mathbf{q}, \mathbf{G},|\mathbf{q}+\mathbf{G}| \neq 0} V(\mathbf{q}+\mathbf{G}) \delta \rho_{\mathbf{q}+\mathbf{G}} \delta \rho_{-\mathbf{q}-\mathbf{G}}
	\end{aligned}
	\end{equation}
where $\epsilon_{m, \tau}(\mathbf{k})$ is the eigenvalue of the continuum model in Eq.~\eqref{eq:eq1}. We define the long-ranged single gate (screened) Coulomb potential: $V(\bq) =\frac{e^{2}}{4 \pi \varepsilon} \int d^{2} \br\left(\frac{1}{\br}-\frac{1}{\sqrt{\br^{2}+d^{2}}}\right) e^{i \bq \cdot \br}=\frac{e^{2}}{2 \varepsilon} \frac{1}{q}\left(1-e^{-q d}\right)$. Here $\frac{d}{2}$ is the distance between graphene layer and single gate, with $d=40$ nm and $\varepsilon=7 \varepsilon_{0}$. The volume $\Omega=N_{\mathbf{k}} \frac{\sqrt{3}}{2} L_{M}^{2}$
with $N_{\mathrm{k}}$ being the number of momentum points in a $\mathrm{mBZ}$ (e.g., $N_{\mathrm{k}}=81$ for a $9 \times 9$ mesh). We choose the bare dispersion, as it is shown in Ref.~\cite{JKang2020RG} that  the renormalization from remote band has been considered in our form of interaction. While it is worth noticing in \cite{Bultinck2020,Bultinck2021,Kwan2021} 
, 
the mean field contribution of remote band interaction from flat band is removed. Whether this remote band interaction is strong enough to change parameter of moir\'e potential obviously is under debate. In our work, we choose the case where flat band approximation is reasonable to carry out our simulation. 

The problem associated with projected Coulomb interaction is solved via a discrete Hubbard-Stratonovich transformation~\cite{Assaad2008,YDLiao2021PRX,YDLiao2019PRL,XuZhang2021}, $e^{\alpha \hat{O}^{2}}=\frac{1}{4} \sum_{l=\pm 1,\pm 2} \gamma(l) e^{\sqrt{\alpha} \eta(l) \hat{o}}+O\left(\alpha^{4}\right)$
(details are shown in the Sec. I of Supplemental Material (SM)~\cite{suppl}). 

{\it Exact ground states in the flat-band limit}\,---\,
When the kinetic energy is ignored (i.e., the flat-band limit), the TBG Hamiltonian at charge neutrality has an emergent  $U(4)$ symmetry  and ground states can be obtained exactly~\cite{Bultinck2020,JYLee2021,bernevig2020tbg5,Vafek2021}. To see the exact solution, one just needs to realize that the valley polarized state, with all electrons in one valley, is a zero-energy eigenstate of $H_{int}$. Because $H_{int}$ is semi-positive definite, this must be a ground state. In addition, any $U(4)$ transformation of this ground state is also a degenerate ground state, including the VP, IVC and spin polarized states, as well as many other degenerate states. For simplicity, in this Letter, we will focus only on the VP and IVC states.

We define the VP and IVC order parameters as $\mathcal{O}_{a}(\boldsymbol{q},\tau)  \equiv \sum_{\boldsymbol{k}} d_{\boldsymbol{k}+\boldsymbol{q}}^{\dagger}(\tau) M_a d_{\boldsymbol{k}}(\tau)$, with $M_a= \tau_z \eta_0$ ($\eta_0$ for band index) for VP and $M_a= \tau_x \eta_y$ or $\tau_y \eta_y$ for the IVC states~\cite{Bultinck2020,KhalafSoftmodes2020,JPLiu2021,bernevig2020tbg5,JYLee2021}. It is straightforward to verify that at $q=0$, these three order parameters obey the commutation relations $[\mathcal{O}_a, \mathcal{O}_b] = i\epsilon_{a,b,c} \mathcal{O}_c$ and they all commute with the interaction Hamiltonian $[O_a,H_{int}]=0$. Thus, they generate a $SU(2)$ symmetry group, a subgroup of the full $U(4)$ symmetry. In the ordered phase, the nonzero expectation value of these order parameters spontaneously breaks this $SU(2)$ symmetry, resulting in spin-wave-like gapless Goldstone modes, i.e. valley waves. Same as ferromagnetism, such valley waves have a quadratic dispersion $\omega \propto k^2$ at low-energy.

As for single-particle excitations, all these degenerate ground states are insulators with a gap proportional to the interaction strength. In the flat-band limit, single-particle Green's function can be calculated exactly at $T=0$~\cite{bernevig2020tbg5}. Despite of the strong Coulomb repulsion, electrons/holes exhibit free-fermion-like behavior, where the Green's function shows four fermion bands with zero damping: two conduction (valence) bands above (below) the Fermi energy.

In a real TBG, away from the flat-band limit, this $SU(2)$ symmetry is explicitly broken by the kinetic energy down to $Z_2$ (valley) and $U_v(1)$ (valley charge conservation), lifting the degeneracy between VP and IVC states. Here, an IVC (VP) state breaks the continuous $U(1)$ (discrete $Z_2$) symmetry, and dynamics fluctuations in VP and IVC states shall exhibit different behaviors. However, if the kinetic energy term is small (i.e., small band width), an approximate $SU(2)$ symmetry may survive, and qualitative features  may still resemble the flat-band limit. The momentum space QMC technique offers a probe to directly visualize the breaking of the $SU(2)$ symmetry as well as the remnant approximate symmetry. 

\begin{figure*}[tb]
\centering
\includegraphics[width=\linewidth]{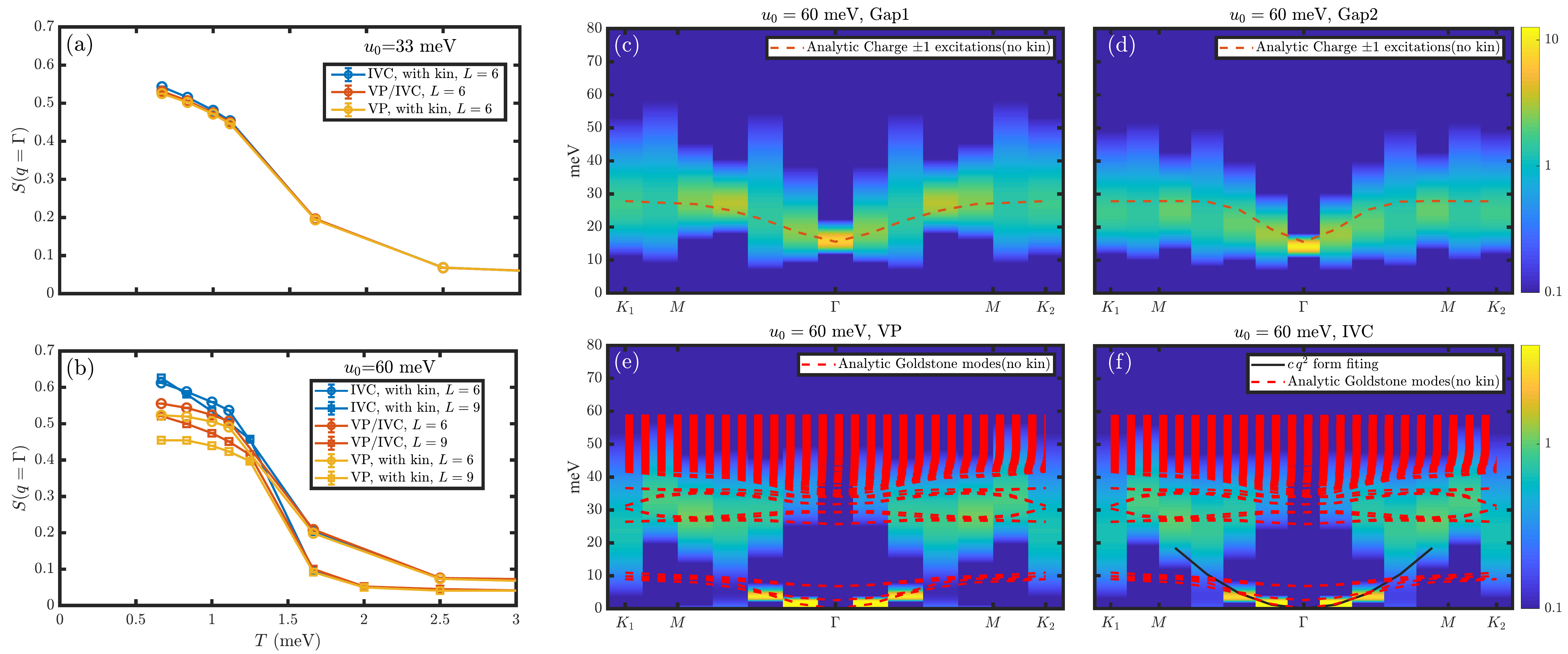}
\caption{(a) $S(\mathbf{q}=\Gamma,\tau=0)$, the squares of order parameters, for VP and IVC at $u_0=33$ meV and $L=6$, as a function of temperature. (b) The same quantity at $u_0=60$ meV with both $L=6$ and 9. When kinetic energy is ignored, the two order parameters are degenerate due to an emergent $SU(2)$ [$U(4)$] symmetry. When the kinetic energy is taken into account ("with kin"), which breaks the symmetry, this degeneracy is lifted. At $u_0=33$ meV, the splitting between VP and IVC is weak. This splitting becomes more pronounced at $u_0=60$ meV,  indicating that IVC is more favored at low temperatures in comparison to VP, although the competition between these two symmetry-breaking channels remains. (c) and (d) single-particle spectra at $T =0.667$ meV, $u_0=60$ meV and $L=9$, which shows an insulating gap $\sim 10$ meV. The dashed lines are the analytic computation of the single-particle dispersion at the flat-band limit following Ref.~\cite{bernevig2020tbg5}. (e) and (f) dynamical spectra of VP and IVC with the same parameters. Sharp and ferromagnetic-like valley waves are observed in both channels near $\mathbf{q}=\Gamma$ and a fit of $c\, q^2$ gives rise to $c=31.32\pm 0.03 \;meV/k_{\theta}^2$ (black solid line in (f)). At the energy scale of twice the single-particle gap, $\sim 20$ meV, valley waves are over-damped into the particle-hole continuum. The dashed lines are the analytic computation of the Goldstone mode at the flat-band limit following Ref.~\cite{bernevig2020tbg5}.}
	\label{fig:fig2}
\end{figure*}

{\it Results and Analysis}\,---\,
In a previous work~\cite{XuZhang2021}, we have shown that $H_{int}$ acquires a correlated insulator ground state at CNP. In this study, we added the kinetic term $H_0$ and carried out the simulations at $u_0=33$ meV and $60$ meV with $6\times6$ and $9\times 9$ momentum meshes. Here $u_0=60$ meV is a realistic case \cite{YiZhang2020,bernevig2020tbg1,song2020tbg2,bernevig2020tbg3,Ashvin2019} which leads to a bandwidth of 1.08 meV. And $u_0=33$ meV is a case between the realistic one and chiral limit. The single-particle spectra are shown in Fig.~\ref{fig:fig1} (b) and (c). The bare (non-interacting) dispersions are depicted as red stars. At low-temperature, for both  $u_0=33$ meV and $60$ meV, interactions push the fermion states away from the Fermi energy, results in an interaction-driven band gap of $\sim 20$ meV, magnitudes larger than that of the bare bandwidth.  Although we are using realistic parameters away from the flat-band limit, as shown in Fig.~\ref{fig:fig2} (c) and (d), the peak of single particle spectra agrees nicely with the solution of the flat-band limit~\cite{bernevig2020tbg5}, indicating that the system is not far from the exactly-solvable limit. As for the width of the peak, due to the finite temperature and the presence of kinetic energy, fermions here exhibit some damping of the order $10$ meV, which is significantly larger than $T$ and the band width of the bare dispersion. This is in contrast to the exactly-solvable limit at $T=0$ where the damping vanishes.

The next question is to reveal the symmetry-breaking channels of this insulating state. The proposed symmetry-breaking states at the CNP, based on Hartree-Fock mean-field analysis, are gradually pointing towards the IVC and VP states~\cite{JPLiu2019Pseudo,JPLiu2021,Bultinck2020,YiZhang2020}. Here, we calculate their corresponding (dynamical) correlation 
\begin{equation}
S_{a}(\boldsymbol{q},\tau) \equiv \frac{1}{N_k^{2}}\left\langle\mathcal{O}_{a}(-\boldsymbol{q},\tau) \mathcal{O}_{a}(\boldsymbol{q},0)\right\rangle
\label{eq:eq6}
\end{equation} 
where $\mathcal{O}_{a}$ is the order parameter of the VP or IVC state defined early on. For static properties, we calculate the equal-time correlation at imaginary time $\tau=0$. To obtain dynamic response, time-dependent $S_{a}(\boldsymbol{q},\tau)$  is calculated at $\tau\in[0,\beta]$, followed by the stochastic analytic continuation (SAC)~\cite{Sandvik1998,beach2004identifying,Olav2008,Sandvik2016,HShao2017,GYSun2018,NSMa2018,li2020kosterlitz,jiang2020,zhou2020amplitude,ZYan2021,ChuangChen2021} to obtain the real frequency spectra~\cite{suppl}.

The static order parameters are presented in Fig.~\ref{fig:fig2} (a) and (b), where we calculate $S(\mathbf{q}=\Gamma,\tau=0)$, the squares of the order parameter, for IVC and VP as a function of temperature. Without the kinetic energy ($H=H_{int}$), IVC and VP share identical susceptibility, which reflects the $SU(2)$ symmetry of the flat-band limit. Once the kinetic energy is included ("with kin" in the Fig.~\ref{fig:fig2} (a) and (b)), this degeneracy is lifted. At $u_0=33$ meV, a small splitting between IVC and VP correlation functions is observed. The splitting becomes more significant when $u_0$ reaches $60$ meV, closer to the realistic case~\cite{Nam2017,Ashvin20192}, with IVC being the more favored ground state. It is worthwhile to note that when system size goes from $6\times6$ to $9\times 9$, the IVC order $S(\mathbf{q}=\Gamma)$ does not change, whereas the VP $S(\mathbf{q}=\Gamma)$ decreases as the system size increases. One shall also notice that although the degeneracy between IVC and VP is lifted, both correlation functions grow at low $T$, indicating that the competition between IVC and VP remains strong and there is no a completely dominant symmetry-breaking channel~\cite{JPLiu2021}.

In addition to static correlations, we also compute the dynamic correlations of IVC and VP as defined in Eq.~\eqref{eq:eq6} and their spectra with the system size of $9\times9$ for the realistic case with kinetic energy at $u_0=60$ meV at low temperature $T=0.667$ meV, much lower than the scale of the single-particle gap. The results are shown in Fig.~\ref{fig:fig2} (e) and (f), with Fig.~\ref{fig:fig2} (c) and (d) the associated single-particle spectra. The dashed lines mark the single-particle dispersion and Goldstone modes when the kinetic energy is ignored~\cite{bernevig2020tbg5}. Measured from $\omega=0$, the single-particle gap is of size $\sim 10$ meV and both the VP and IVC spectra develop a clear and sharp valley wave dispersion at low-energy near $\Gamma$. Remarkably, although the static susceptibility indicates that the $SU(2)$ symmetry has been explicitly broken at $u_0=60$ meV and the degeneracy between IVC and VP is lifted [Fig.~\ref{fig:fig2} (b)],  the IVC and VP spectra are almost identical and are strikingly similar to the flat-band limit~\cite{bernevig2020tbg5,Feldner2011}. These sharp Goldstone-like modes are in strong analog to $SU(2)$ ferromagnetic Goldstone modes with $\omega \propto c\,q^2$ and $c=31.32\pm 0.03 \;meV/k_{\theta}^2$, (where $k_{\theta}=8 \pi \sin (\theta / 2) /\left(3 a_{0}\right)$ and the lattice constant of the monolayer graphene $a_{0}=0.246 \mathrm{~nm}$), indicating an approximate $SU(2)$ symmetry survives in our model. It is worthwhile to highlight that this SU(2) approximate symmetry is not an exact symmetry and it breaks at low energy. Thus, at very small $q$ and $\omega$, this magnon-like excitation will exhibit a linear dispersion $\omega \propto q$, due to the broken SU(2) symmetry \cite{KhalafSoftmodes2020}. For our study, because this SU(2) symmetry breaking is really weak, such linear dispersion is not visible in the QMC data.

One other interesting feature of these valley waves is that above the energy scale of $\sim 20$ meV, the sharp collective excitations become heavily damped, which is not seen in analytical solution(dashed line in Fig.~\ref{fig:fig2} (e) and (f)). The analytical solutions(without kinetic energy) are only consistent with QMC results(with kinetic energy) at low energy mode near $\Gamma$ point means that our results are beyond the mean-field type of calculations.  The damping of collective modes has two origins (1) scattering between collective modes and (2) damping due to the fermion particle-hole continuum. The second damping channel arises for energy larger than twice of the fermion gap, and thus is responsible for the over-damped features at energy above $20$ meV shown in Fig.~\ref{fig:fig2} (e) and (f). This is in strong analogy to the damping of ferromagnetic spin excitations in the graphene nanoribbons, where the flat band gives rise to the ferromagnetic long-range order but the spin waves becomes over-damped in the particle-hole continuum~\cite{Feldner2011,Golor2013,Golor2014}. 

{\it Discussion and outlook}\,---\, Quantum dynamics of collective excitations holds the key to the understanding of many-body effects in twisted bilayer graphene and other quantum moir\'e systems. This study suggests that the momentum-space QMC method offers a powerful tool to tackle this problem. In particular, the spectral function obtained via this unbiased method offers a bridge way to directly connect theoretical studies with experimental measurements, especially spectroscopy methods, such as inelastic light- or neutron- scattering and tunneling spectroscopy, making it possible to compare measurements in experiments and large-scale quantum simulations at the quantitative level.

\begin{acknowledgments}
{\it Acknowledgments}\,---\, We are indebted to Yi Zhang for the help in the form factor tables. We thank Tianyu Qiao, Jian Kang, Jianpeng Liu and Xi Dai for stimulating discussions.G.P.P., X.Z. and Z.Y.M. acknowledge support from the RGC of Hong Kong SAR of China (Grant Nos. 17303019, 17301420, 17301721 and AoE/P701/20), the Strategic Priority Research Program of the Chinese Academy of Sciences (Grant No. XDB33000000), the K. C. Wong Education Foundation (Grant No. GJTD-2020-01) and the Seed Funding “QuantumInspired explainable-AI” at the HKU-TCL Joint Research Centre for Artificial Intelligence. H.L. and K.S. acknowledge support through NSF Grant No.NSF-EFMA-1741618. We thank the Computational Initiative at the Faculty of Science and the Information Technology Services at the University of Hong Kong and the Tianhe platforms at the National Supercomputer Center in Guangzhou for their technical support and generous allocation of CPU time.
\end{acknowledgments}

\bibliography{TBG_IVC.bib}

\begin{thebibliography}{73}%
\makeatletter
\providecommand \@ifxundefined [1]{%
 \@ifx{#1\undefined}
}%
\providecommand \@ifnum [1]{%
 \ifnum #1\expandafter \@firstoftwo
 \else \expandafter \@secondoftwo
 \fi
}%
\providecommand \@ifx [1]{%
 \ifx #1\expandafter \@firstoftwo
 \else \expandafter \@secondoftwo
 \fi
}%
\providecommand \natexlab [1]{#1}%
\providecommand \enquote  [1]{``#1''}%
\providecommand \bibnamefont  [1]{#1}%
\providecommand \bibfnamefont [1]{#1}%
\providecommand \citenamefont [1]{#1}%
\providecommand \href@noop [0]{\@secondoftwo}%
\providecommand \href [0]{\begingroup \@sanitize@url \@href}%
\providecommand \@href[1]{\@@startlink{#1}\@@href}%
\providecommand \@@href[1]{\endgroup#1\@@endlink}%
\providecommand \@sanitize@url [0]{\catcode `\\12\catcode `\$12\catcode
  `\&12\catcode `\#12\catcode `\^12\catcode `\_12\catcode `\%12\relax}%
\providecommand \@@startlink[1]{}%
\providecommand \@@endlink[0]{}%
\providecommand \url  [0]{\begingroup\@sanitize@url \@url }%
\providecommand \@url [1]{\endgroup\@href {#1}{\urlprefix }}%
\providecommand \urlprefix  [0]{URL }%
\providecommand \Eprint [0]{\href }%
\providecommand \doibase [0]{https://doi.org/}%
\providecommand \selectlanguage [0]{\@gobble}%
\providecommand \bibinfo  [0]{\@secondoftwo}%
\providecommand \bibfield  [0]{\@secondoftwo}%
\providecommand \translation [1]{[#1]}%
\providecommand \BibitemOpen [0]{}%
\providecommand \bibitemStop [0]{}%
\providecommand \bibitemNoStop [0]{.\EOS\space}%
\providecommand \EOS [0]{\spacefactor3000\relax}%
\providecommand \BibitemShut  [1]{\csname bibitem#1\endcsname}%
\let\auto@bib@innerbib\@empty
\bibitem [{\citenamefont {Trambly~de Laissardière}\ \emph
  {et~al.}(2010)\citenamefont {Trambly~de Laissardière}, \citenamefont
  {Mayou},\ and\ \citenamefont {Magaud}}]{Trambly2010}%
  \BibitemOpen
  \bibfield  {author} {\bibinfo {author} {\bibfnamefont {G.}~\bibnamefont
  {Trambly~de Laissardière}}, \bibinfo {author} {\bibfnamefont
  {D.}~\bibnamefont {Mayou}},\ and\ \bibinfo {author} {\bibfnamefont
  {L.}~\bibnamefont {Magaud}},\ }\bibfield  {title} {\bibinfo {title}
  {Localization of dirac electrons in rotated graphene bilayers},\ }\href
  {https://doi.org/doi: 10.1021/nl902948m} {\bibfield  {journal} {\bibinfo
  {journal} {Nano Letters}\ }\textbf {\bibinfo {volume} {10}},\ \bibinfo
  {pages} {804 } (\bibinfo {year} {2010})}\BibitemShut {NoStop}%
\bibitem [{\citenamefont {Trambly~de Laissardi\`ere}\ \emph
  {et~al.}(2012)\citenamefont {Trambly~de Laissardi\`ere}, \citenamefont
  {Mayou},\ and\ \citenamefont {Magaud}}]{Trambly2012}%
  \BibitemOpen
  \bibfield  {author} {\bibinfo {author} {\bibfnamefont {G.}~\bibnamefont
  {Trambly~de Laissardi\`ere}}, \bibinfo {author} {\bibfnamefont
  {D.}~\bibnamefont {Mayou}},\ and\ \bibinfo {author} {\bibfnamefont
  {L.}~\bibnamefont {Magaud}},\ }\bibfield  {title} {\bibinfo {title}
  {Numerical studies of confined states in rotated bilayers of graphene},\
  }\href {https://doi.org/10.1103/PhysRevB.86.125413} {\bibfield  {journal}
  {\bibinfo  {journal} {Phys. Rev. B}\ }\textbf {\bibinfo {volume} {86}},\
  \bibinfo {pages} {125413} (\bibinfo {year} {2012})}\BibitemShut {NoStop}%
\bibitem [{\citenamefont {Bistritzer}\ and\ \citenamefont
  {MacDonald}(2011)}]{Bistritzer_TBG}%
  \BibitemOpen
  \bibfield  {author} {\bibinfo {author} {\bibfnamefont {R.}~\bibnamefont
  {Bistritzer}}\ and\ \bibinfo {author} {\bibfnamefont {A.~H.}\ \bibnamefont
  {MacDonald}},\ }\bibfield  {title} {\bibinfo {title} {Moir{\'e} bands in
  twisted double-layer graphene},\ }\href
  {https://doi.org/10.1073/pnas.1108174108} {\bibfield  {journal} {\bibinfo
  {journal} {Proceedings of the National Academy of Sciences}\ }\textbf
  {\bibinfo {volume} {108}},\ \bibinfo {pages} {12233} (\bibinfo {year}
  {2011})}\BibitemShut {NoStop}%
\bibitem [{\citenamefont {Rozhkov}\ \emph {et~al.}(2016)\citenamefont
  {Rozhkov}, \citenamefont {Sboychakov}, \citenamefont {Rakhmanov},\ and\
  \citenamefont {Nori}}]{ROZHKOV20161}%
  \BibitemOpen
  \bibfield  {author} {\bibinfo {author} {\bibfnamefont {A.}~\bibnamefont
  {Rozhkov}}, \bibinfo {author} {\bibfnamefont {A.}~\bibnamefont {Sboychakov}},
  \bibinfo {author} {\bibfnamefont {A.}~\bibnamefont {Rakhmanov}},\ and\
  \bibinfo {author} {\bibfnamefont {F.}~\bibnamefont {Nori}},\ }\bibfield
  {title} {\bibinfo {title} {Electronic properties of graphene-based bilayer
  systems},\ }\href
  {https://doi.org/https://doi.org/10.1016/j.physrep.2016.07.003} {\bibfield
  {journal} {\bibinfo  {journal} {Physics Reports}\ }\textbf {\bibinfo {volume}
  {648}},\ \bibinfo {pages} {1} (\bibinfo {year} {2016})},\ \bibinfo {note}
  {electronic properties of graphene-based bilayer systems}\BibitemShut
  {NoStop}%
\bibitem [{\citenamefont {Lopes~dos Santos}\ \emph {et~al.}(2007)\citenamefont
  {Lopes~dos Santos}, \citenamefont {Peres},\ and\ \citenamefont
  {Castro~Neto}}]{Santos2007}%
  \BibitemOpen
  \bibfield  {author} {\bibinfo {author} {\bibfnamefont {J.~M.~B.}\
  \bibnamefont {Lopes~dos Santos}}, \bibinfo {author} {\bibfnamefont
  {N.~M.~R.}\ \bibnamefont {Peres}},\ and\ \bibinfo {author} {\bibfnamefont
  {A.~H.}\ \bibnamefont {Castro~Neto}},\ }\bibfield  {title} {\bibinfo {title}
  {Graphene bilayer with a twist: Electronic structure},\ }\href
  {https://doi.org/10.1103/PhysRevLett.99.256802} {\bibfield  {journal}
  {\bibinfo  {journal} {Phys. Rev. Lett.}\ }\textbf {\bibinfo {volume} {99}},\
  \bibinfo {pages} {256802} (\bibinfo {year} {2007})}\BibitemShut {NoStop}%
\bibitem [{\citenamefont {Lopes~dos Santos}\ \emph {et~al.}(2012)\citenamefont
  {Lopes~dos Santos}, \citenamefont {Peres},\ and\ \citenamefont
  {Castro~Neto}}]{Santos2012}%
  \BibitemOpen
  \bibfield  {author} {\bibinfo {author} {\bibfnamefont {J.~M.~B.}\
  \bibnamefont {Lopes~dos Santos}}, \bibinfo {author} {\bibfnamefont
  {N.~M.~R.}\ \bibnamefont {Peres}},\ and\ \bibinfo {author} {\bibfnamefont
  {A.~H.}\ \bibnamefont {Castro~Neto}},\ }\bibfield  {title} {\bibinfo {title}
  {Continuum model of the twisted graphene bilayer},\ }\href
  {https://doi.org/10.1103/PhysRevB.86.155449} {\bibfield  {journal} {\bibinfo
  {journal} {Phys. Rev. B}\ }\textbf {\bibinfo {volume} {86}},\ \bibinfo
  {pages} {155449} (\bibinfo {year} {2012})}\BibitemShut {NoStop}%
\bibitem [{\citenamefont {Cao}\ \emph {et~al.}(2018{\natexlab{a}})\citenamefont
  {Cao}, \citenamefont {Fatemi}, \citenamefont {Fang}, \citenamefont
  {Watanabe}, \citenamefont {Taniguchi}, \citenamefont {Kaxiras},\ and\
  \citenamefont {Jarillo-Herrero}}]{cao2018unconventional}%
  \BibitemOpen
  \bibfield  {author} {\bibinfo {author} {\bibfnamefont {Y.}~\bibnamefont
  {Cao}}, \bibinfo {author} {\bibfnamefont {V.}~\bibnamefont {Fatemi}},
  \bibinfo {author} {\bibfnamefont {S.}~\bibnamefont {Fang}}, \bibinfo {author}
  {\bibfnamefont {K.}~\bibnamefont {Watanabe}}, \bibinfo {author}
  {\bibfnamefont {T.}~\bibnamefont {Taniguchi}}, \bibinfo {author}
  {\bibfnamefont {E.}~\bibnamefont {Kaxiras}},\ and\ \bibinfo {author}
  {\bibfnamefont {P.}~\bibnamefont {Jarillo-Herrero}},\ }\bibfield  {title}
  {\bibinfo {title} {Unconventional superconductivity in magic-angle graphene
  superlattices},\ }\href {https://doi.org/10.1038/nature26160} {\bibfield
  {journal} {\bibinfo  {journal} {Nature}\ }\textbf {\bibinfo {volume} {556}},\
  \bibinfo {pages} {43} (\bibinfo {year} {2018}{\natexlab{a}})}\BibitemShut
  {NoStop}%
\bibitem [{\citenamefont {Cao}\ \emph {et~al.}(2018{\natexlab{b}})\citenamefont
  {Cao}, \citenamefont {Fatemi}, \citenamefont {Demir}, \citenamefont {Fang},
  \citenamefont {Tomarken}, \citenamefont {Luo}, \citenamefont
  {Sanchez-Yamagishi}, \citenamefont {Watanabe}, \citenamefont {Taniguchi},
  \citenamefont {Kaxiras} \emph {et~al.}}]{cao2018correlated}%
  \BibitemOpen
  \bibfield  {author} {\bibinfo {author} {\bibfnamefont {Y.}~\bibnamefont
  {Cao}}, \bibinfo {author} {\bibfnamefont {V.}~\bibnamefont {Fatemi}},
  \bibinfo {author} {\bibfnamefont {A.}~\bibnamefont {Demir}}, \bibinfo
  {author} {\bibfnamefont {S.}~\bibnamefont {Fang}}, \bibinfo {author}
  {\bibfnamefont {S.~L.}\ \bibnamefont {Tomarken}}, \bibinfo {author}
  {\bibfnamefont {J.~Y.}\ \bibnamefont {Luo}}, \bibinfo {author} {\bibfnamefont
  {J.~D.}\ \bibnamefont {Sanchez-Yamagishi}}, \bibinfo {author} {\bibfnamefont
  {K.}~\bibnamefont {Watanabe}}, \bibinfo {author} {\bibfnamefont
  {T.}~\bibnamefont {Taniguchi}}, \bibinfo {author} {\bibfnamefont
  {E.}~\bibnamefont {Kaxiras}}, \emph {et~al.},\ }\bibfield  {title} {\bibinfo
  {title} {Correlated insulator behaviour at half-filling in magic-angle
  graphene superlattices},\ }\href {https://doi.org/10.1038/nature26154}
  {\bibfield  {journal} {\bibinfo  {journal} {Nature}\ }\textbf {\bibinfo
  {volume} {556}},\ \bibinfo {pages} {80} (\bibinfo {year}
  {2018}{\natexlab{b}})}\BibitemShut {NoStop}%
\bibitem [{\citenamefont {Chen}\ \emph {et~al.}(2020)\citenamefont {Chen},
  \citenamefont {Sharpe}, \citenamefont {Fox}, \citenamefont {Zhang},
  \citenamefont {Wang}, \citenamefont {Jiang}, \citenamefont {Lyu},
  \citenamefont {Li}, \citenamefont {Watanabe}, \citenamefont {Taniguchi} \emph
  {et~al.}}]{chen2020tunable}%
  \BibitemOpen
  \bibfield  {author} {\bibinfo {author} {\bibfnamefont {G.}~\bibnamefont
  {Chen}}, \bibinfo {author} {\bibfnamefont {A.~L.}\ \bibnamefont {Sharpe}},
  \bibinfo {author} {\bibfnamefont {E.~J.}\ \bibnamefont {Fox}}, \bibinfo
  {author} {\bibfnamefont {Y.-H.}\ \bibnamefont {Zhang}}, \bibinfo {author}
  {\bibfnamefont {S.}~\bibnamefont {Wang}}, \bibinfo {author} {\bibfnamefont
  {L.}~\bibnamefont {Jiang}}, \bibinfo {author} {\bibfnamefont
  {B.}~\bibnamefont {Lyu}}, \bibinfo {author} {\bibfnamefont {H.}~\bibnamefont
  {Li}}, \bibinfo {author} {\bibfnamefont {K.}~\bibnamefont {Watanabe}},
  \bibinfo {author} {\bibfnamefont {T.}~\bibnamefont {Taniguchi}}, \emph
  {et~al.},\ }\bibfield  {title} {\bibinfo {title} {Tunable correlated chern
  insulator and ferromagnetism in a moir{\'e} superlattice},\ }\href
  {https://doi.org/10.1038/s41586-020-2049-7} {\bibfield  {journal} {\bibinfo
  {journal} {Nature}\ }\textbf {\bibinfo {volume} {579}},\ \bibinfo {pages}
  {56} (\bibinfo {year} {2020})}\BibitemShut {NoStop}%
\bibitem [{\citenamefont {Kerelsky}\ \emph {et~al.}(2019)\citenamefont
  {Kerelsky}, \citenamefont {McGilly}, \citenamefont {Kennes}, \citenamefont
  {Xian}, \citenamefont {Yankowitz}, \citenamefont {Chen}, \citenamefont
  {Watanabe}, \citenamefont {Taniguchi}, \citenamefont {Hone}, \citenamefont
  {Dean} \emph {et~al.}}]{kerelsky2019maximized}%
  \BibitemOpen
  \bibfield  {author} {\bibinfo {author} {\bibfnamefont {A.}~\bibnamefont
  {Kerelsky}}, \bibinfo {author} {\bibfnamefont {L.~J.}\ \bibnamefont
  {McGilly}}, \bibinfo {author} {\bibfnamefont {D.~M.}\ \bibnamefont {Kennes}},
  \bibinfo {author} {\bibfnamefont {L.}~\bibnamefont {Xian}}, \bibinfo {author}
  {\bibfnamefont {M.}~\bibnamefont {Yankowitz}}, \bibinfo {author}
  {\bibfnamefont {S.}~\bibnamefont {Chen}}, \bibinfo {author} {\bibfnamefont
  {K.}~\bibnamefont {Watanabe}}, \bibinfo {author} {\bibfnamefont
  {T.}~\bibnamefont {Taniguchi}}, \bibinfo {author} {\bibfnamefont
  {J.}~\bibnamefont {Hone}}, \bibinfo {author} {\bibfnamefont {C.}~\bibnamefont
  {Dean}}, \emph {et~al.},\ }\bibfield  {title} {\bibinfo {title} {Maximized
  electron interactions at the magic angle in twisted bilayer graphene},\
  }\href {https://doi.org/10.1038/s41586-019-1431-9} {\bibfield  {journal}
  {\bibinfo  {journal} {Nature}\ }\textbf {\bibinfo {volume} {572}},\ \bibinfo
  {pages} {95} (\bibinfo {year} {2019})}\BibitemShut {NoStop}%
\bibitem [{\citenamefont {Tomarken}\ \emph {et~al.}(2019)\citenamefont
  {Tomarken}, \citenamefont {Cao}, \citenamefont {Demir}, \citenamefont
  {Watanabe}, \citenamefont {Taniguchi}, \citenamefont {Jarillo-Herrero},\ and\
  \citenamefont {Ashoori}}]{Tomarken2019}%
  \BibitemOpen
  \bibfield  {author} {\bibinfo {author} {\bibfnamefont {S.~L.}\ \bibnamefont
  {Tomarken}}, \bibinfo {author} {\bibfnamefont {Y.}~\bibnamefont {Cao}},
  \bibinfo {author} {\bibfnamefont {A.}~\bibnamefont {Demir}}, \bibinfo
  {author} {\bibfnamefont {K.}~\bibnamefont {Watanabe}}, \bibinfo {author}
  {\bibfnamefont {T.}~\bibnamefont {Taniguchi}}, \bibinfo {author}
  {\bibfnamefont {P.}~\bibnamefont {Jarillo-Herrero}},\ and\ \bibinfo {author}
  {\bibfnamefont {R.~C.}\ \bibnamefont {Ashoori}},\ }\bibfield  {title}
  {\bibinfo {title} {Electronic compressibility of magic-angle graphene
  superlattices},\ }\href {https://doi.org/10.1103/PhysRevLett.123.046601}
  {\bibfield  {journal} {\bibinfo  {journal} {Phys. Rev. Lett.}\ }\textbf
  {\bibinfo {volume} {123}},\ \bibinfo {pages} {046601} (\bibinfo {year}
  {2019})}\BibitemShut {NoStop}%
\bibitem [{\citenamefont {Lu}\ \emph {et~al.}(2019)\citenamefont {Lu},
  \citenamefont {Stepanov}, \citenamefont {Yang}, \citenamefont {Xie},
  \citenamefont {Aamir}, \citenamefont {Das}, \citenamefont {Urgell},
  \citenamefont {Watanabe}, \citenamefont {Taniguchi}, \citenamefont {Zhang}
  \emph {et~al.}}]{lu2019superconductors}%
  \BibitemOpen
  \bibfield  {author} {\bibinfo {author} {\bibfnamefont {X.}~\bibnamefont
  {Lu}}, \bibinfo {author} {\bibfnamefont {P.}~\bibnamefont {Stepanov}},
  \bibinfo {author} {\bibfnamefont {W.}~\bibnamefont {Yang}}, \bibinfo {author}
  {\bibfnamefont {M.}~\bibnamefont {Xie}}, \bibinfo {author} {\bibfnamefont
  {M.~A.}\ \bibnamefont {Aamir}}, \bibinfo {author} {\bibfnamefont
  {I.}~\bibnamefont {Das}}, \bibinfo {author} {\bibfnamefont {C.}~\bibnamefont
  {Urgell}}, \bibinfo {author} {\bibfnamefont {K.}~\bibnamefont {Watanabe}},
  \bibinfo {author} {\bibfnamefont {T.}~\bibnamefont {Taniguchi}}, \bibinfo
  {author} {\bibfnamefont {G.}~\bibnamefont {Zhang}}, \emph {et~al.},\
  }\bibfield  {title} {\bibinfo {title} {Superconductors, orbital magnets and
  correlated states in magic-angle bilayer graphene},\ }\href
  {https://doi.org/10.1038/s41586-019-1695-0} {\bibfield  {journal} {\bibinfo
  {journal} {Nature}\ }\textbf {\bibinfo {volume} {574}},\ \bibinfo {pages}
  {653} (\bibinfo {year} {2019})}\BibitemShut {NoStop}%
\bibitem [{\citenamefont {Xie}\ \emph {et~al.}(2019)\citenamefont {Xie},
  \citenamefont {Lian}, \citenamefont {J{\"a}ck}, \citenamefont {Liu},
  \citenamefont {Chiu}, \citenamefont {Watanabe}, \citenamefont {Taniguchi},
  \citenamefont {Bernevig},\ and\ \citenamefont
  {Yazdani}}]{xie2019spectroscopic}%
  \BibitemOpen
  \bibfield  {author} {\bibinfo {author} {\bibfnamefont {Y.}~\bibnamefont
  {Xie}}, \bibinfo {author} {\bibfnamefont {B.}~\bibnamefont {Lian}}, \bibinfo
  {author} {\bibfnamefont {B.}~\bibnamefont {J{\"a}ck}}, \bibinfo {author}
  {\bibfnamefont {X.}~\bibnamefont {Liu}}, \bibinfo {author} {\bibfnamefont
  {C.-L.}\ \bibnamefont {Chiu}}, \bibinfo {author} {\bibfnamefont
  {K.}~\bibnamefont {Watanabe}}, \bibinfo {author} {\bibfnamefont
  {T.}~\bibnamefont {Taniguchi}}, \bibinfo {author} {\bibfnamefont {B.~A.}\
  \bibnamefont {Bernevig}},\ and\ \bibinfo {author} {\bibfnamefont
  {A.}~\bibnamefont {Yazdani}},\ }\bibfield  {title} {\bibinfo {title}
  {Spectroscopic signatures of many-body correlations in magic-angle twisted
  bilayer graphene},\ }\href {https://doi.org/10.1038/s41586-019-1422-x}
  {\bibfield  {journal} {\bibinfo  {journal} {Nature}\ }\textbf {\bibinfo
  {volume} {572}},\ \bibinfo {pages} {101} (\bibinfo {year}
  {2019})}\BibitemShut {NoStop}%
\bibitem [{\citenamefont {Shen}\ \emph {et~al.}(2020)\citenamefont {Shen},
  \citenamefont {Chu}, \citenamefont {Wu}, \citenamefont {Li}, \citenamefont
  {Wang}, \citenamefont {Zhao}, \citenamefont {Tang}, \citenamefont {Liu},
  \citenamefont {Tian}, \citenamefont {Watanabe}, \citenamefont {Taniguchi},
  \citenamefont {Yang}, \citenamefont {Meng}, \citenamefont {Shi},
  \citenamefont {Yazyev},\ and\ \citenamefont {Zhang}}]{Shen2020DTBG}%
  \BibitemOpen
  \bibfield  {author} {\bibinfo {author} {\bibfnamefont {C.}~\bibnamefont
  {Shen}}, \bibinfo {author} {\bibfnamefont {Y.}~\bibnamefont {Chu}}, \bibinfo
  {author} {\bibfnamefont {Q.}~\bibnamefont {Wu}}, \bibinfo {author}
  {\bibfnamefont {N.}~\bibnamefont {Li}}, \bibinfo {author} {\bibfnamefont
  {S.}~\bibnamefont {Wang}}, \bibinfo {author} {\bibfnamefont {Y.}~\bibnamefont
  {Zhao}}, \bibinfo {author} {\bibfnamefont {J.}~\bibnamefont {Tang}}, \bibinfo
  {author} {\bibfnamefont {J.}~\bibnamefont {Liu}}, \bibinfo {author}
  {\bibfnamefont {J.}~\bibnamefont {Tian}}, \bibinfo {author} {\bibfnamefont
  {K.}~\bibnamefont {Watanabe}}, \bibinfo {author} {\bibfnamefont
  {T.}~\bibnamefont {Taniguchi}}, \bibinfo {author} {\bibfnamefont
  {R.}~\bibnamefont {Yang}}, \bibinfo {author} {\bibfnamefont {Z.~Y.}\
  \bibnamefont {Meng}}, \bibinfo {author} {\bibfnamefont {D.}~\bibnamefont
  {Shi}}, \bibinfo {author} {\bibfnamefont {O.~V.}\ \bibnamefont {Yazyev}},\
  and\ \bibinfo {author} {\bibfnamefont {G.}~\bibnamefont {Zhang}},\ }\bibfield
   {title} {\bibinfo {title} {Correlated states in twisted double bilayer
  graphene},\ }\bibfield  {journal} {\bibinfo  {journal} {Nature Physics}\
  }\href {https://doi.org/10.1038/s41567-020-0825-9}
  {10.1038/s41567-020-0825-9} (\bibinfo {year} {2020})\BibitemShut {NoStop}%
\bibitem [{\citenamefont {Nuckolls}\ \emph {et~al.}(2020)\citenamefont
  {Nuckolls}, \citenamefont {Oh}, \citenamefont {Wong}, \citenamefont {Lian},
  \citenamefont {Watanabe}, \citenamefont {Taniguchi}, \citenamefont
  {Bernevig},\ and\ \citenamefont {Yazdani}}]{Nuckolls_2020}%
  \BibitemOpen
  \bibfield  {author} {\bibinfo {author} {\bibfnamefont {K.~P.}\ \bibnamefont
  {Nuckolls}}, \bibinfo {author} {\bibfnamefont {M.}~\bibnamefont {Oh}},
  \bibinfo {author} {\bibfnamefont {D.}~\bibnamefont {Wong}}, \bibinfo {author}
  {\bibfnamefont {B.}~\bibnamefont {Lian}}, \bibinfo {author} {\bibfnamefont
  {K.}~\bibnamefont {Watanabe}}, \bibinfo {author} {\bibfnamefont
  {T.}~\bibnamefont {Taniguchi}}, \bibinfo {author} {\bibfnamefont {B.~A.}\
  \bibnamefont {Bernevig}},\ and\ \bibinfo {author} {\bibfnamefont
  {A.}~\bibnamefont {Yazdani}},\ }\bibfield  {title} {\bibinfo {title}
  {Strongly correlated chern insulators in magic-angle twisted bilayer
  graphene},\ }\href {https://doi.org/10.1038/s41586-020-3028-8} {\bibfield
  {journal} {\bibinfo  {journal} {Nature}\ }\textbf {\bibinfo {volume} {588}},\
  \bibinfo {pages} {610} (\bibinfo {year} {2020})}\BibitemShut {NoStop}%
\bibitem [{\citenamefont {Pierce}\ \emph {et~al.}(2021)\citenamefont {Pierce},
  \citenamefont {Xie}, \citenamefont {Park}, \citenamefont {Khalaf},
  \citenamefont {Lee}, \citenamefont {Cao}, \citenamefont {Parker},
  \citenamefont {Forrester}, \citenamefont {Chen}, \citenamefont {Watanabe},
  \citenamefont {Taniguchi}, \citenamefont {Vishwanath}, \citenamefont
  {Jarillo-Herrero},\ and\ \citenamefont {Yacoby}}]{pierce2021unconventional}%
  \BibitemOpen
  \bibfield  {author} {\bibinfo {author} {\bibfnamefont {A.~T.}\ \bibnamefont
  {Pierce}}, \bibinfo {author} {\bibfnamefont {Y.}~\bibnamefont {Xie}},
  \bibinfo {author} {\bibfnamefont {J.~M.}\ \bibnamefont {Park}}, \bibinfo
  {author} {\bibfnamefont {E.}~\bibnamefont {Khalaf}}, \bibinfo {author}
  {\bibfnamefont {S.~H.}\ \bibnamefont {Lee}}, \bibinfo {author} {\bibfnamefont
  {Y.}~\bibnamefont {Cao}}, \bibinfo {author} {\bibfnamefont {D.~E.}\
  \bibnamefont {Parker}}, \bibinfo {author} {\bibfnamefont {P.~R.}\
  \bibnamefont {Forrester}}, \bibinfo {author} {\bibfnamefont {S.}~\bibnamefont
  {Chen}}, \bibinfo {author} {\bibfnamefont {K.}~\bibnamefont {Watanabe}},
  \bibinfo {author} {\bibfnamefont {T.}~\bibnamefont {Taniguchi}}, \bibinfo
  {author} {\bibfnamefont {A.}~\bibnamefont {Vishwanath}}, \bibinfo {author}
  {\bibfnamefont {P.}~\bibnamefont {Jarillo-Herrero}},\ and\ \bibinfo {author}
  {\bibfnamefont {A.}~\bibnamefont {Yacoby}},\ }\href
  {https://arxiv.org/abs/2101.04123} {\bibinfo {title} {Unconventional sequence
  of correlated chern insulators in magic-angle twisted bilayer graphene}}
  (\bibinfo {year} {2021}),\ \Eprint {https://arxiv.org/abs/2101.04123}
  {arXiv:2101.04123 [cond-mat.mes-hall]} \BibitemShut {NoStop}%
\bibitem [{\citenamefont {{Moriyama}}\ \emph {et~al.}(2019)\citenamefont
  {{Moriyama}}, \citenamefont {{Morita}}, \citenamefont {{Komatsu}},
  \citenamefont {{Endo}}, \citenamefont {{Iwasaki}}, \citenamefont
  {{Nakaharai}}, \citenamefont {{Noguchi}}, \citenamefont {{Wakayama}},
  \citenamefont {{Watanabe}}, \citenamefont {{Tsuya}}, \citenamefont
  {{Watanabe}},\ and\ \citenamefont {{Taniguchi}}}]{moriyama2019}%
  \BibitemOpen
  \bibfield  {author} {\bibinfo {author} {\bibfnamefont {S.}~\bibnamefont
  {{Moriyama}}}, \bibinfo {author} {\bibfnamefont {Y.}~\bibnamefont
  {{Morita}}}, \bibinfo {author} {\bibfnamefont {K.}~\bibnamefont {{Komatsu}}},
  \bibinfo {author} {\bibfnamefont {K.}~\bibnamefont {{Endo}}}, \bibinfo
  {author} {\bibfnamefont {T.}~\bibnamefont {{Iwasaki}}}, \bibinfo {author}
  {\bibfnamefont {S.}~\bibnamefont {{Nakaharai}}}, \bibinfo {author}
  {\bibfnamefont {Y.}~\bibnamefont {{Noguchi}}}, \bibinfo {author}
  {\bibfnamefont {Y.}~\bibnamefont {{Wakayama}}}, \bibinfo {author}
  {\bibfnamefont {E.}~\bibnamefont {{Watanabe}}}, \bibinfo {author}
  {\bibfnamefont {D.}~\bibnamefont {{Tsuya}}}, \bibinfo {author} {\bibfnamefont
  {K.}~\bibnamefont {{Watanabe}}},\ and\ \bibinfo {author} {\bibfnamefont
  {T.}~\bibnamefont {{Taniguchi}}},\ }\bibfield  {title} {\bibinfo {title}
  {{Observation of superconductivity in bilayer graphene/hexagonal boron
  nitride superlattices}},\ }\href@noop {} {\bibfield  {journal} {\bibinfo
  {journal} {arXiv e-prints}\ ,\ \bibinfo {eid} {arXiv:1901.09356}} (\bibinfo
  {year} {2019})},\ \Eprint {https://arxiv.org/abs/1901.09356}
  {arXiv:1901.09356 [cond-mat.supr-con]} \BibitemShut {NoStop}%
\bibitem [{\citenamefont {Rozen}\ \emph {et~al.}(2021)\citenamefont {Rozen},
  \citenamefont {Park}, \citenamefont {Zondiner}, \citenamefont {Cao},
  \citenamefont {Rodan-Legrain}, \citenamefont {Taniguchi}, \citenamefont
  {Watanabe}, \citenamefont {Oreg}, \citenamefont {Stern}, \citenamefont
  {Berg}, \citenamefont {Jarillo-Herrero},\ and\ \citenamefont
  {Ilani}}]{Rozen2021}%
  \BibitemOpen
  \bibfield  {author} {\bibinfo {author} {\bibfnamefont {A.}~\bibnamefont
  {Rozen}}, \bibinfo {author} {\bibfnamefont {J.~M.}\ \bibnamefont {Park}},
  \bibinfo {author} {\bibfnamefont {U.}~\bibnamefont {Zondiner}}, \bibinfo
  {author} {\bibfnamefont {Y.}~\bibnamefont {Cao}}, \bibinfo {author}
  {\bibfnamefont {D.}~\bibnamefont {Rodan-Legrain}}, \bibinfo {author}
  {\bibfnamefont {T.}~\bibnamefont {Taniguchi}}, \bibinfo {author}
  {\bibfnamefont {K.}~\bibnamefont {Watanabe}}, \bibinfo {author}
  {\bibfnamefont {Y.}~\bibnamefont {Oreg}}, \bibinfo {author} {\bibfnamefont
  {A.}~\bibnamefont {Stern}}, \bibinfo {author} {\bibfnamefont
  {E.}~\bibnamefont {Berg}}, \bibinfo {author} {\bibfnamefont {P.}~\bibnamefont
  {Jarillo-Herrero}},\ and\ \bibinfo {author} {\bibfnamefont {S.}~\bibnamefont
  {Ilani}},\ }\bibfield  {title} {\bibinfo {title} {Entropic evidence for a
  pomeranchuk effect in magic-angle graphene},\ }\href
  {https://doi.org/10.1038/s41586-021-03319-3} {\bibfield  {journal} {\bibinfo
  {journal} {Nature}\ }\textbf {\bibinfo {volume} {592}},\ \bibinfo {pages}
  {214 } (\bibinfo {year} {2021})}\BibitemShut {NoStop}%
\bibitem [{\citenamefont {Liu}\ \emph {et~al.}(2020)\citenamefont {Liu},
  \citenamefont {Chiu}, \citenamefont {Lee}, \citenamefont {Farahi},
  \citenamefont {Watanabe}, \citenamefont {Taniguchi}, \citenamefont
  {Vishwanath},\ and\ \citenamefont {Yazdani}}]{liu2020spectroscopy}%
  \BibitemOpen
  \bibfield  {author} {\bibinfo {author} {\bibfnamefont {X.}~\bibnamefont
  {Liu}}, \bibinfo {author} {\bibfnamefont {C.-L.}\ \bibnamefont {Chiu}},
  \bibinfo {author} {\bibfnamefont {J.~Y.}\ \bibnamefont {Lee}}, \bibinfo
  {author} {\bibfnamefont {G.}~\bibnamefont {Farahi}}, \bibinfo {author}
  {\bibfnamefont {K.}~\bibnamefont {Watanabe}}, \bibinfo {author}
  {\bibfnamefont {T.}~\bibnamefont {Taniguchi}}, \bibinfo {author}
  {\bibfnamefont {A.}~\bibnamefont {Vishwanath}},\ and\ \bibinfo {author}
  {\bibfnamefont {A.}~\bibnamefont {Yazdani}},\ }\bibfield  {title} {\bibinfo
  {title} {Spectroscopy of a tunable moir$\backslash$'e system with a
  correlated and topological flat band},\ }\href
  {https://arxiv.org/abs/2008.07552} {\bibfield  {journal} {\bibinfo  {journal}
  {arXiv preprint arXiv:2008.07552}\ } (\bibinfo {year} {2020})}\BibitemShut
  {NoStop}%
\bibitem [{\citenamefont {{Khalaf}}\ \emph {et~al.}(2020)\citenamefont
  {{Khalaf}}, \citenamefont {{Bultinck}}, \citenamefont {{Vishwanath}},\ and\
  \citenamefont {{Zaletel}}}]{KhalafSoftmodes2020}%
  \BibitemOpen
  \bibfield  {author} {\bibinfo {author} {\bibfnamefont {E.}~\bibnamefont
  {{Khalaf}}}, \bibinfo {author} {\bibfnamefont {N.}~\bibnamefont
  {{Bultinck}}}, \bibinfo {author} {\bibfnamefont {A.}~\bibnamefont
  {{Vishwanath}}},\ and\ \bibinfo {author} {\bibfnamefont {M.~P.}\ \bibnamefont
  {{Zaletel}}},\ }\bibfield  {title} {\bibinfo {title} {{Soft modes in magic
  angle twisted bilayer graphene}},\ }\href@noop {} {\bibfield  {journal}
  {\bibinfo  {journal} {arXiv e-prints}\ ,\ \bibinfo {eid} {arXiv:2009.14827}}
  (\bibinfo {year} {2020})},\ \Eprint {https://arxiv.org/abs/2009.14827}
  {arXiv:2009.14827 [cond-mat.str-el]} \BibitemShut {NoStop}%
\bibitem [{\citenamefont {Soejima}\ \emph {et~al.}(2020)\citenamefont
  {Soejima}, \citenamefont {Parker}, \citenamefont {Bultinck}, \citenamefont
  {Hauschild},\ and\ \citenamefont {Zaletel}}]{Zaletel2020}%
  \BibitemOpen
  \bibfield  {author} {\bibinfo {author} {\bibfnamefont {T.}~\bibnamefont
  {Soejima}}, \bibinfo {author} {\bibfnamefont {D.~E.}\ \bibnamefont {Parker}},
  \bibinfo {author} {\bibfnamefont {N.}~\bibnamefont {Bultinck}}, \bibinfo
  {author} {\bibfnamefont {J.}~\bibnamefont {Hauschild}},\ and\ \bibinfo
  {author} {\bibfnamefont {M.~P.}\ \bibnamefont {Zaletel}},\ }\bibfield
  {title} {\bibinfo {title} {Efficient simulation of moir\'e materials using
  the density matrix renormalization group},\ }\href
  {https://doi.org/10.1103/PhysRevB.102.205111} {\bibfield  {journal} {\bibinfo
   {journal} {Phys. Rev. B}\ }\textbf {\bibinfo {volume} {102}},\ \bibinfo
  {pages} {205111} (\bibinfo {year} {2020})}\BibitemShut {NoStop}%
\bibitem [{\citenamefont {Khalaf}\ \emph {et~al.}(2021)\citenamefont {Khalaf},
  \citenamefont {Chatterjee}, \citenamefont {Bultinck}, \citenamefont
  {Zaletel},\ and\ \citenamefont {Vishwanath}}]{Khalaf2021}%
  \BibitemOpen
  \bibfield  {author} {\bibinfo {author} {\bibfnamefont {E.}~\bibnamefont
  {Khalaf}}, \bibinfo {author} {\bibfnamefont {S.}~\bibnamefont {Chatterjee}},
  \bibinfo {author} {\bibfnamefont {N.}~\bibnamefont {Bultinck}}, \bibinfo
  {author} {\bibfnamefont {M.~P.}\ \bibnamefont {Zaletel}},\ and\ \bibinfo
  {author} {\bibfnamefont {A.}~\bibnamefont {Vishwanath}},\ }\bibfield  {title}
  {\bibinfo {title} {Charged skyrmions and topological origin of
  superconductivity in magic-angle graphene},\ }\bibfield  {journal} {\bibinfo
  {journal} {Science Advances}\ }\textbf {\bibinfo {volume} {7}},\ \href
  {https://doi.org/10.1126/sciadv.abf5299} {10.1126/sciadv.abf5299} (\bibinfo
  {year} {2021})\BibitemShut {NoStop}%
\bibitem [{\citenamefont {{Chatterjee}}\ \emph {et~al.}(2020)\citenamefont
  {{Chatterjee}}, \citenamefont {{Ippoliti}},\ and\ \citenamefont
  {{Zaletel}}}]{Chatterjee2020}%
  \BibitemOpen
  \bibfield  {author} {\bibinfo {author} {\bibfnamefont {S.}~\bibnamefont
  {{Chatterjee}}}, \bibinfo {author} {\bibfnamefont {M.}~\bibnamefont
  {{Ippoliti}}},\ and\ \bibinfo {author} {\bibfnamefont {M.~P.}\ \bibnamefont
  {{Zaletel}}},\ }\bibfield  {title} {\bibinfo {title} {{Skyrmion
  Superconductivity: DMRG evidence for a topological route to
  superconductivity}},\ }\href {https://arxiv.org/abs/2010.01144} {\bibfield
  {journal} {\bibinfo  {journal} {arXiv e-prints}\ ,\ \bibinfo {eid}
  {arXiv:2010.01144}} (\bibinfo {year} {2020})},\ \Eprint
  {https://arxiv.org/abs/2010.01144} {arXiv:2010.01144 [cond-mat.str-el]}
  \BibitemShut {NoStop}%
\bibitem [{\citenamefont {Koshino}\ \emph {et~al.}(2018)\citenamefont
  {Koshino}, \citenamefont {Yuan}, \citenamefont {Koretsune}, \citenamefont
  {Ochi}, \citenamefont {Kuroki},\ and\ \citenamefont {Fu}}]{Koshino2018}%
  \BibitemOpen
  \bibfield  {author} {\bibinfo {author} {\bibfnamefont {M.}~\bibnamefont
  {Koshino}}, \bibinfo {author} {\bibfnamefont {N.~F.~Q.}\ \bibnamefont
  {Yuan}}, \bibinfo {author} {\bibfnamefont {T.}~\bibnamefont {Koretsune}},
  \bibinfo {author} {\bibfnamefont {M.}~\bibnamefont {Ochi}}, \bibinfo {author}
  {\bibfnamefont {K.}~\bibnamefont {Kuroki}},\ and\ \bibinfo {author}
  {\bibfnamefont {L.}~\bibnamefont {Fu}},\ }\bibfield  {title} {\bibinfo
  {title} {Maximally localized wannier orbitals and the extended hubbard model
  for twisted bilayer graphene},\ }\href
  {https://doi.org/10.1103/PhysRevX.8.031087} {\bibfield  {journal} {\bibinfo
  {journal} {Phys. Rev. X}\ }\textbf {\bibinfo {volume} {8}},\ \bibinfo {pages}
  {031087} (\bibinfo {year} {2018})}\BibitemShut {NoStop}%
\bibitem [{\citenamefont {Kang}\ and\ \citenamefont {Vafek}(2018)}]{Kang2018}%
  \BibitemOpen
  \bibfield  {author} {\bibinfo {author} {\bibfnamefont {J.}~\bibnamefont
  {Kang}}\ and\ \bibinfo {author} {\bibfnamefont {O.}~\bibnamefont {Vafek}},\
  }\bibfield  {title} {\bibinfo {title} {Symmetry, maximally localized wannier
  states, and a low-energy model for twisted bilayer graphene narrow bands},\
  }\href {https://doi.org/10.1103/PhysRevX.8.031088} {\bibfield  {journal}
  {\bibinfo  {journal} {Phys. Rev. X}\ }\textbf {\bibinfo {volume} {8}},\
  \bibinfo {pages} {031088} (\bibinfo {year} {2018})}\BibitemShut {NoStop}%
\bibitem [{\citenamefont {Xu}\ \emph {et~al.}(2018)\citenamefont {Xu},
  \citenamefont {Law},\ and\ \citenamefont {Lee}}]{XYXu2018}%
  \BibitemOpen
  \bibfield  {author} {\bibinfo {author} {\bibfnamefont {X.~Y.}\ \bibnamefont
  {Xu}}, \bibinfo {author} {\bibfnamefont {K.~T.}\ \bibnamefont {Law}},\ and\
  \bibinfo {author} {\bibfnamefont {P.~A.}\ \bibnamefont {Lee}},\ }\bibfield
  {title} {\bibinfo {title} {Kekul\'e valence bond order in an extended hubbard
  model on the honeycomb lattice with possible applications to twisted bilayer
  graphene},\ }\href {https://doi.org/10.1103/PhysRevB.98.121406} {\bibfield
  {journal} {\bibinfo  {journal} {Phys. Rev. B}\ }\textbf {\bibinfo {volume}
  {98}},\ \bibinfo {pages} {121406} (\bibinfo {year} {2018})}\BibitemShut
  {NoStop}%
\bibitem [{\citenamefont {Kang}\ and\ \citenamefont {Vafek}(2019)}]{Kang2019}%
  \BibitemOpen
  \bibfield  {author} {\bibinfo {author} {\bibfnamefont {J.}~\bibnamefont
  {Kang}}\ and\ \bibinfo {author} {\bibfnamefont {O.}~\bibnamefont {Vafek}},\
  }\bibfield  {title} {\bibinfo {title} {Strong coupling phases of partially
  filled twisted bilayer graphene narrow bands},\ }\href
  {https://doi.org/10.1103/PhysRevLett.122.246401} {\bibfield  {journal}
  {\bibinfo  {journal} {Phys. Rev. Lett.}\ }\textbf {\bibinfo {volume} {122}},\
  \bibinfo {pages} {246401} (\bibinfo {year} {2019})}\BibitemShut {NoStop}%
\bibitem [{\citenamefont {Liao}\ \emph
  {et~al.}(2021{\natexlab{a}})\citenamefont {Liao}, \citenamefont {Kang},
  \citenamefont {Brei\o{}}, \citenamefont {Xu}, \citenamefont {Wu},
  \citenamefont {Andersen}, \citenamefont {Fernandes},\ and\ \citenamefont
  {Meng}}]{YDLiao2021PRX}%
  \BibitemOpen
  \bibfield  {author} {\bibinfo {author} {\bibfnamefont {Y.~D.}\ \bibnamefont
  {Liao}}, \bibinfo {author} {\bibfnamefont {J.}~\bibnamefont {Kang}}, \bibinfo
  {author} {\bibfnamefont {C.~N.}\ \bibnamefont {Brei\o{}}}, \bibinfo {author}
  {\bibfnamefont {X.~Y.}\ \bibnamefont {Xu}}, \bibinfo {author} {\bibfnamefont
  {H.-Q.}\ \bibnamefont {Wu}}, \bibinfo {author} {\bibfnamefont {B.~M.}\
  \bibnamefont {Andersen}}, \bibinfo {author} {\bibfnamefont {R.~M.}\
  \bibnamefont {Fernandes}},\ and\ \bibinfo {author} {\bibfnamefont {Z.~Y.}\
  \bibnamefont {Meng}},\ }\bibfield  {title} {\bibinfo {title}
  {Correlation-induced insulating topological phases at charge neutrality in
  twisted bilayer graphene},\ }\href
  {https://doi.org/10.1103/PhysRevX.11.011014} {\bibfield  {journal} {\bibinfo
  {journal} {Phys. Rev. X}\ }\textbf {\bibinfo {volume} {11}},\ \bibinfo
  {pages} {011014} (\bibinfo {year} {2021}{\natexlab{a}})}\BibitemShut
  {NoStop}%
\bibitem [{\citenamefont {Liao}\ \emph
  {et~al.}(2021{\natexlab{b}})\citenamefont {Liao}, \citenamefont {Xu},
  \citenamefont {Meng},\ and\ \citenamefont {Kang}}]{YDLiao2021CPB}%
  \BibitemOpen
  \bibfield  {author} {\bibinfo {author} {\bibfnamefont {Y.-D.}\ \bibnamefont
  {Liao}}, \bibinfo {author} {\bibfnamefont {X.-Y.}\ \bibnamefont {Xu}},
  \bibinfo {author} {\bibfnamefont {Z.-Y.}\ \bibnamefont {Meng}},\ and\
  \bibinfo {author} {\bibfnamefont {J.}~\bibnamefont {Kang}},\ }\bibfield
  {title} {\bibinfo {title} {Correlated insulating phases in the twisted
  bilayer graphene},\ }\href {https://doi.org/10.1088/1674-1056/abcfa3}
  {\bibfield  {journal} {\bibinfo  {journal} {Chinese Physics B}\ }\textbf
  {\bibinfo {volume} {30}},\ \bibinfo {pages} {017305} (\bibinfo {year}
  {2021}{\natexlab{b}})}\BibitemShut {NoStop}%
\bibitem [{\citenamefont {Chen}\ \emph
  {et~al.}(2021{\natexlab{a}})\citenamefont {Chen}, \citenamefont {Liao},
  \citenamefont {Chen}, \citenamefont {Vafek}, \citenamefont {Kang},
  \citenamefont {Li},\ and\ \citenamefont {Meng}}]{BBChen2020}%
  \BibitemOpen
  \bibfield  {author} {\bibinfo {author} {\bibfnamefont {B.-B.}\ \bibnamefont
  {Chen}}, \bibinfo {author} {\bibfnamefont {Y.~D.}\ \bibnamefont {Liao}},
  \bibinfo {author} {\bibfnamefont {Z.}~\bibnamefont {Chen}}, \bibinfo {author}
  {\bibfnamefont {O.}~\bibnamefont {Vafek}}, \bibinfo {author} {\bibfnamefont
  {J.}~\bibnamefont {Kang}}, \bibinfo {author} {\bibfnamefont {W.}~\bibnamefont
  {Li}},\ and\ \bibinfo {author} {\bibfnamefont {Z.~Y.}\ \bibnamefont {Meng}},\
  }\bibfield  {title} {\bibinfo {title} {Realization of topological mott
  insulator in a twisted bilayer graphene lattice model},\ }\href
  {https://doi.org/10.1038/s41467-021-25438-1} {\bibfield  {journal} {\bibinfo
  {journal} {Nature Communications}\ }\textbf {\bibinfo {volume} {12}},\
  \bibinfo {pages} {5480} (\bibinfo {year} {2021}{\natexlab{a}})}\BibitemShut
  {NoStop}%
\bibitem [{\citenamefont {Po}\ \emph {et~al.}(2018{\natexlab{a}})\citenamefont
  {Po}, \citenamefont {Zou}, \citenamefont {Vishwanath},\ and\ \citenamefont
  {Senthil}}]{Ashvin2018}%
  \BibitemOpen
  \bibfield  {author} {\bibinfo {author} {\bibfnamefont {H.~C.}\ \bibnamefont
  {Po}}, \bibinfo {author} {\bibfnamefont {L.}~\bibnamefont {Zou}}, \bibinfo
  {author} {\bibfnamefont {A.}~\bibnamefont {Vishwanath}},\ and\ \bibinfo
  {author} {\bibfnamefont {T.}~\bibnamefont {Senthil}},\ }\bibfield  {title}
  {\bibinfo {title} {Origin of mott insulating behavior and superconductivity
  in twisted bilayer graphene},\ }\href
  {https://doi.org/10.1103/PhysRevX.8.031089} {\bibfield  {journal} {\bibinfo
  {journal} {Phys. Rev. X}\ }\textbf {\bibinfo {volume} {8}},\ \bibinfo {pages}
  {031089} (\bibinfo {year} {2018}{\natexlab{a}})}\BibitemShut {NoStop}%
\bibitem [{\citenamefont {Po}\ \emph {et~al.}(2018{\natexlab{b}})\citenamefont
  {Po}, \citenamefont {Watanabe},\ and\ \citenamefont
  {Vishwanath}}]{HCPo2018Fragile}%
  \BibitemOpen
  \bibfield  {author} {\bibinfo {author} {\bibfnamefont {H.~C.}\ \bibnamefont
  {Po}}, \bibinfo {author} {\bibfnamefont {H.}~\bibnamefont {Watanabe}},\ and\
  \bibinfo {author} {\bibfnamefont {A.}~\bibnamefont {Vishwanath}},\ }\bibfield
   {title} {\bibinfo {title} {Fragile topology and wannier obstructions},\
  }\href {https://doi.org/10.1103/PhysRevLett.121.126402} {\bibfield  {journal}
  {\bibinfo  {journal} {Phys. Rev. Lett.}\ }\textbf {\bibinfo {volume} {121}},\
  \bibinfo {pages} {126402} (\bibinfo {year} {2018}{\natexlab{b}})}\BibitemShut
  {NoStop}%
\bibitem [{\citenamefont {Po}\ \emph {et~al.}(2019)\citenamefont {Po},
  \citenamefont {Zou}, \citenamefont {Senthil},\ and\ \citenamefont
  {Vishwanath}}]{HCPo2019}%
  \BibitemOpen
  \bibfield  {author} {\bibinfo {author} {\bibfnamefont {H.~C.}\ \bibnamefont
  {Po}}, \bibinfo {author} {\bibfnamefont {L.}~\bibnamefont {Zou}}, \bibinfo
  {author} {\bibfnamefont {T.}~\bibnamefont {Senthil}},\ and\ \bibinfo {author}
  {\bibfnamefont {A.}~\bibnamefont {Vishwanath}},\ }\bibfield  {title}
  {\bibinfo {title} {Faithful tight-binding models and fragile topology of
  magic-angle bilayer graphene},\ }\href
  {https://doi.org/10.1103/PhysRevB.99.195455} {\bibfield  {journal} {\bibinfo
  {journal} {Phys. Rev. B}\ }\textbf {\bibinfo {volume} {99}},\ \bibinfo
  {pages} {195455} (\bibinfo {year} {2019})}\BibitemShut {NoStop}%
\bibitem [{\citenamefont {Bernevig}\ \emph
  {et~al.}(2021{\natexlab{a}})\citenamefont {Bernevig}, \citenamefont {Lian},
  \citenamefont {Cowsik}, \citenamefont {Xie}, \citenamefont {Regnault},\ and\
  \citenamefont {Song}}]{bernevig2020tbg5}%
  \BibitemOpen
  \bibfield  {author} {\bibinfo {author} {\bibfnamefont {B.~A.}\ \bibnamefont
  {Bernevig}}, \bibinfo {author} {\bibfnamefont {B.}~\bibnamefont {Lian}},
  \bibinfo {author} {\bibfnamefont {A.}~\bibnamefont {Cowsik}}, \bibinfo
  {author} {\bibfnamefont {F.}~\bibnamefont {Xie}}, \bibinfo {author}
  {\bibfnamefont {N.}~\bibnamefont {Regnault}},\ and\ \bibinfo {author}
  {\bibfnamefont {Z.-D.}\ \bibnamefont {Song}},\ }\bibfield  {title} {\bibinfo
  {title} {Twisted bilayer graphene. v. exact analytic many-body excitations in
  coulomb hamiltonians: Charge gap, goldstone modes, and absence of cooper
  pairing},\ }\href {https://doi.org/10.1103/PhysRevB.103.205415} {\bibfield
  {journal} {\bibinfo  {journal} {Phys. Rev. B}\ }\textbf {\bibinfo {volume}
  {103}},\ \bibinfo {pages} {205415} (\bibinfo {year}
  {2021}{\natexlab{a}})}\BibitemShut {NoStop}%
\bibitem [{\citenamefont {Bultinck}\ \emph {et~al.}(2020)\citenamefont
  {Bultinck}, \citenamefont {Khalaf}, \citenamefont {Liu}, \citenamefont
  {Chatterjee}, \citenamefont {Vishwanath},\ and\ \citenamefont
  {Zaletel}}]{Bultinck2020}%
  \BibitemOpen
  \bibfield  {author} {\bibinfo {author} {\bibfnamefont {N.}~\bibnamefont
  {Bultinck}}, \bibinfo {author} {\bibfnamefont {E.}~\bibnamefont {Khalaf}},
  \bibinfo {author} {\bibfnamefont {S.}~\bibnamefont {Liu}}, \bibinfo {author}
  {\bibfnamefont {S.}~\bibnamefont {Chatterjee}}, \bibinfo {author}
  {\bibfnamefont {A.}~\bibnamefont {Vishwanath}},\ and\ \bibinfo {author}
  {\bibfnamefont {M.~P.}\ \bibnamefont {Zaletel}},\ }\bibfield  {title}
  {\bibinfo {title} {Ground state and hidden symmetry of magic-angle graphene
  at even integer filling},\ }\href
  {https://doi.org/10.1103/PhysRevX.10.031034} {\bibfield  {journal} {\bibinfo
  {journal} {Phys. Rev. X}\ }\textbf {\bibinfo {volume} {10}},\ \bibinfo
  {pages} {031034} (\bibinfo {year} {2020})}\BibitemShut {NoStop}%
\bibitem [{\citenamefont {Zhang}\ \emph {et~al.}(2020)\citenamefont {Zhang},
  \citenamefont {Jiang}, \citenamefont {Wang},\ and\ \citenamefont
  {Zhang}}]{YiZhang2020}%
  \BibitemOpen
  \bibfield  {author} {\bibinfo {author} {\bibfnamefont {Y.}~\bibnamefont
  {Zhang}}, \bibinfo {author} {\bibfnamefont {K.}~\bibnamefont {Jiang}},
  \bibinfo {author} {\bibfnamefont {Z.}~\bibnamefont {Wang}},\ and\ \bibinfo
  {author} {\bibfnamefont {F.}~\bibnamefont {Zhang}},\ }\bibfield  {title}
  {\bibinfo {title} {Correlated insulating phases of twisted bilayer graphene
  at commensurate filling fractions: A hartree-fock study},\ }\href
  {https://doi.org/10.1103/PhysRevB.102.035136} {\bibfield  {journal} {\bibinfo
   {journal} {Phys. Rev. B}\ }\textbf {\bibinfo {volume} {102}},\ \bibinfo
  {pages} {035136} (\bibinfo {year} {2020})}\BibitemShut {NoStop}%
\bibitem [{\citenamefont {Vafek}\ and\ \citenamefont
  {Kang}(2020)}]{JKang2020RG}%
  \BibitemOpen
  \bibfield  {author} {\bibinfo {author} {\bibfnamefont {O.}~\bibnamefont
  {Vafek}}\ and\ \bibinfo {author} {\bibfnamefont {J.}~\bibnamefont {Kang}},\
  }\bibfield  {title} {\bibinfo {title} {Renormalization group study of hidden
  symmetry in twisted bilayer graphene with coulomb interactions},\ }\href
  {https://doi.org/10.1103/PhysRevLett.125.257602} {\bibfield  {journal}
  {\bibinfo  {journal} {Phys. Rev. Lett.}\ }\textbf {\bibinfo {volume} {125}},\
  \bibinfo {pages} {257602} (\bibinfo {year} {2020})}\BibitemShut {NoStop}%
\bibitem [{\citenamefont {Parker}\ \emph {et~al.}(2021)\citenamefont {Parker},
  \citenamefont {Soejima}, \citenamefont {Hauschild}, \citenamefont {Zaletel},\
  and\ \citenamefont {Bultinck}}]{Bultinck2021}%
  \BibitemOpen
  \bibfield  {author} {\bibinfo {author} {\bibfnamefont {D.~E.}\ \bibnamefont
  {Parker}}, \bibinfo {author} {\bibfnamefont {T.}~\bibnamefont {Soejima}},
  \bibinfo {author} {\bibfnamefont {J.}~\bibnamefont {Hauschild}}, \bibinfo
  {author} {\bibfnamefont {M.~P.}\ \bibnamefont {Zaletel}},\ and\ \bibinfo
  {author} {\bibfnamefont {N.}~\bibnamefont {Bultinck}},\ }\bibfield  {title}
  {\bibinfo {title} {Strain-induced quantum phase transitions in magic-angle
  graphene},\ }\href {https://doi.org/10.1103/PhysRevLett.127.027601}
  {\bibfield  {journal} {\bibinfo  {journal} {Phys. Rev. Lett.}\ }\textbf
  {\bibinfo {volume} {127}},\ \bibinfo {pages} {027601} (\bibinfo {year}
  {2021})}\BibitemShut {NoStop}%
\bibitem [{\citenamefont {Liu}\ \emph {et~al.}(2019{\natexlab{a}})\citenamefont
  {Liu}, \citenamefont {Liu},\ and\ \citenamefont {Dai}}]{JPLiu2019Pseudo}%
  \BibitemOpen
  \bibfield  {author} {\bibinfo {author} {\bibfnamefont {J.}~\bibnamefont
  {Liu}}, \bibinfo {author} {\bibfnamefont {J.}~\bibnamefont {Liu}},\ and\
  \bibinfo {author} {\bibfnamefont {X.}~\bibnamefont {Dai}},\ }\bibfield
  {title} {\bibinfo {title} {Pseudo landau level representation of twisted
  bilayer graphene: Band topology and implications on the correlated insulating
  phase},\ }\href {https://doi.org/10.1103/PhysRevB.99.155415} {\bibfield
  {journal} {\bibinfo  {journal} {Phys. Rev. B}\ }\textbf {\bibinfo {volume}
  {99}},\ \bibinfo {pages} {155415} (\bibinfo {year}
  {2019}{\natexlab{a}})}\BibitemShut {NoStop}%
\bibitem [{\citenamefont {Liu}\ and\ \citenamefont {Dai}(2021)}]{JPLiu2021}%
  \BibitemOpen
  \bibfield  {author} {\bibinfo {author} {\bibfnamefont {J.}~\bibnamefont
  {Liu}}\ and\ \bibinfo {author} {\bibfnamefont {X.}~\bibnamefont {Dai}},\
  }\bibfield  {title} {\bibinfo {title} {Theories for the correlated insulating
  states and quantum anomalous hall effect phenomena in twisted bilayer
  graphene},\ }\href {https://doi.org/10.1103/PhysRevB.103.035427} {\bibfield
  {journal} {\bibinfo  {journal} {Phys. Rev. B}\ }\textbf {\bibinfo {volume}
  {103}},\ \bibinfo {pages} {035427} (\bibinfo {year} {2021})}\BibitemShut
  {NoStop}%
\bibitem [{\citenamefont {Lian}\ \emph {et~al.}(2021)\citenamefont {Lian},
  \citenamefont {Song}, \citenamefont {Regnault}, \citenamefont {Efetov},
  \citenamefont {Yazdani},\ and\ \citenamefont {Bernevig}}]{lian2020tbg4}%
  \BibitemOpen
  \bibfield  {author} {\bibinfo {author} {\bibfnamefont {B.}~\bibnamefont
  {Lian}}, \bibinfo {author} {\bibfnamefont {Z.-D.}\ \bibnamefont {Song}},
  \bibinfo {author} {\bibfnamefont {N.}~\bibnamefont {Regnault}}, \bibinfo
  {author} {\bibfnamefont {D.~K.}\ \bibnamefont {Efetov}}, \bibinfo {author}
  {\bibfnamefont {A.}~\bibnamefont {Yazdani}},\ and\ \bibinfo {author}
  {\bibfnamefont {B.~A.}\ \bibnamefont {Bernevig}},\ }\bibfield  {title}
  {\bibinfo {title} {Twisted bilayer graphene. iv. exact insulator ground
  states and phase diagram},\ }\href
  {https://doi.org/10.1103/PhysRevB.103.205414} {\bibfield  {journal} {\bibinfo
   {journal} {Phys. Rev. B}\ }\textbf {\bibinfo {volume} {103}},\ \bibinfo
  {pages} {205414} (\bibinfo {year} {2021})}\BibitemShut {NoStop}%
\bibitem [{\citenamefont {{Kwan}}\ \emph {et~al.}(2021)\citenamefont {{Kwan}},
  \citenamefont {{Wagner}}, \citenamefont {{Soejima}}, \citenamefont
  {{Zaletel}}, \citenamefont {{Simon}}, \citenamefont {{Parameswaran}},\ and\
  \citenamefont {{Bultinck}}}]{Kwan2021}%
  \BibitemOpen
  \bibfield  {author} {\bibinfo {author} {\bibfnamefont {Y.~H.}\ \bibnamefont
  {{Kwan}}}, \bibinfo {author} {\bibfnamefont {G.}~\bibnamefont {{Wagner}}},
  \bibinfo {author} {\bibfnamefont {T.}~\bibnamefont {{Soejima}}}, \bibinfo
  {author} {\bibfnamefont {M.~P.}\ \bibnamefont {{Zaletel}}}, \bibinfo {author}
  {\bibfnamefont {S.~H.}\ \bibnamefont {{Simon}}}, \bibinfo {author}
  {\bibfnamefont {S.~A.}\ \bibnamefont {{Parameswaran}}},\ and\ \bibinfo
  {author} {\bibfnamefont {N.}~\bibnamefont {{Bultinck}}},\ }\bibfield  {title}
  {\bibinfo {title} {{Kekul{\'e} spiral order at all nonzero integer fillings
  in twisted bilayer graphene}},\ }\href@noop {} {\bibfield  {journal}
  {\bibinfo  {journal} {arXiv e-prints}\ ,\ \bibinfo {eid} {arXiv:2105.05857}}
  (\bibinfo {year} {2021})},\ \Eprint {https://arxiv.org/abs/2105.05857}
  {arXiv:2105.05857 [cond-mat.str-el]} \BibitemShut {NoStop}%
\bibitem [{\citenamefont {Ippoliti}\ \emph {et~al.}(2018)\citenamefont
  {Ippoliti}, \citenamefont {Mong}, \citenamefont {Assaad},\ and\ \citenamefont
  {Zaletel}}]{Ippoliti2018}%
  \BibitemOpen
  \bibfield  {author} {\bibinfo {author} {\bibfnamefont {M.}~\bibnamefont
  {Ippoliti}}, \bibinfo {author} {\bibfnamefont {R.~S.~K.}\ \bibnamefont
  {Mong}}, \bibinfo {author} {\bibfnamefont {F.~F.}\ \bibnamefont {Assaad}},\
  and\ \bibinfo {author} {\bibfnamefont {M.~P.}\ \bibnamefont {Zaletel}},\
  }\bibfield  {title} {\bibinfo {title} {Half-filled landau levels: A continuum
  and sign-free regularization for three-dimensional quantum critical points},\
  }\href {https://doi.org/10.1103/PhysRevB.98.235108} {\bibfield  {journal}
  {\bibinfo  {journal} {Phys. Rev. B}\ }\textbf {\bibinfo {volume} {98}},\
  \bibinfo {pages} {235108} (\bibinfo {year} {2018})}\BibitemShut {NoStop}%
\bibitem [{\citenamefont {Liu}\ \emph {et~al.}(2019{\natexlab{b}})\citenamefont
  {Liu}, \citenamefont {Xu}, \citenamefont {Qi}, \citenamefont {Sun},\ and\
  \citenamefont {Meng}}]{ZHLiuEMUS2019}%
  \BibitemOpen
  \bibfield  {author} {\bibinfo {author} {\bibfnamefont {Z.~H.}\ \bibnamefont
  {Liu}}, \bibinfo {author} {\bibfnamefont {X.~Y.}\ \bibnamefont {Xu}},
  \bibinfo {author} {\bibfnamefont {Y.}~\bibnamefont {Qi}}, \bibinfo {author}
  {\bibfnamefont {K.}~\bibnamefont {Sun}},\ and\ \bibinfo {author}
  {\bibfnamefont {Z.~Y.}\ \bibnamefont {Meng}},\ }\bibfield  {title} {\bibinfo
  {title} {Elective-momentum ultrasize quantum monte carlo method},\ }\href
  {https://doi.org/10.1103/PhysRevB.99.085114} {\bibfield  {journal} {\bibinfo
  {journal} {Phys. Rev. B}\ }\textbf {\bibinfo {volume} {99}},\ \bibinfo
  {pages} {085114} (\bibinfo {year} {2019}{\natexlab{b}})}\BibitemShut
  {NoStop}%
\bibitem [{\citenamefont {Zhang}\ \emph {et~al.}(2021)\citenamefont {Zhang},
  \citenamefont {Pan}, \citenamefont {Zhang}, \citenamefont {Kang},\ and\
  \citenamefont {Meng}}]{XuZhang2021}%
  \BibitemOpen
  \bibfield  {author} {\bibinfo {author} {\bibfnamefont {X.}~\bibnamefont
  {Zhang}}, \bibinfo {author} {\bibfnamefont {G.}~\bibnamefont {Pan}}, \bibinfo
  {author} {\bibfnamefont {Y.}~\bibnamefont {Zhang}}, \bibinfo {author}
  {\bibfnamefont {J.}~\bibnamefont {Kang}},\ and\ \bibinfo {author}
  {\bibfnamefont {Z.~Y.}\ \bibnamefont {Meng}},\ }\bibfield  {title} {\bibinfo
  {title} {Momentum space quantum monte carlo on twisted bilayer graphene},\
  }\href {https://doi.org/10.1088/0256-307X/38/7/077305} {\bibfield  {journal}
  {\bibinfo  {journal} {Chinese Physics Letters}\ }\textbf {\bibinfo {volume}
  {38}},\ \bibinfo {eid} {077305} (\bibinfo {year} {2021})}\BibitemShut
  {NoStop}%
\bibitem [{\citenamefont {{Hofmann}}\ \emph {et~al.}(2021)\citenamefont
  {{Hofmann}}, \citenamefont {{Khalaf}}, \citenamefont {{Vishwanath}},
  \citenamefont {{Berg}},\ and\ \citenamefont {{Lee}}}]{JYLee2021}%
  \BibitemOpen
  \bibfield  {author} {\bibinfo {author} {\bibfnamefont {J.~S.}\ \bibnamefont
  {{Hofmann}}}, \bibinfo {author} {\bibfnamefont {E.}~\bibnamefont {{Khalaf}}},
  \bibinfo {author} {\bibfnamefont {A.}~\bibnamefont {{Vishwanath}}}, \bibinfo
  {author} {\bibfnamefont {E.}~\bibnamefont {{Berg}}},\ and\ \bibinfo {author}
  {\bibfnamefont {J.~Y.}\ \bibnamefont {{Lee}}},\ }\bibfield  {title} {\bibinfo
  {title} {{Fermionic Monte Carlo study of a realistic model of twisted bilayer
  graphene}},\ }\href@noop {} {\bibfield  {journal} {\bibinfo  {journal} {arXiv
  e-prints}\ ,\ \bibinfo {eid} {arXiv:2105.12112}} (\bibinfo {year} {2021})},\
  \Eprint {https://arxiv.org/abs/2105.12112} {arXiv:2105.12112
  [cond-mat.str-el]} \BibitemShut {NoStop}%
\bibitem [{\citenamefont {Sandvik}(2016)}]{Sandvik2016}%
  \BibitemOpen
  \bibfield  {author} {\bibinfo {author} {\bibfnamefont {A.~W.}\ \bibnamefont
  {Sandvik}},\ }\bibfield  {title} {\bibinfo {title} {Constrained sampling
  method for analytic continuation},\ }\href
  {https://doi.org/10.1103/PhysRevE.94.063308} {\bibfield  {journal} {\bibinfo
  {journal} {Phys. Rev. E}\ }\textbf {\bibinfo {volume} {94}},\ \bibinfo
  {pages} {063308} (\bibinfo {year} {2016})}\BibitemShut {NoStop}%
\bibitem [{\citenamefont {Shao}\ \emph {et~al.}(2017)\citenamefont {Shao},
  \citenamefont {Qin}, \citenamefont {Capponi}, \citenamefont {Chesi},
  \citenamefont {Meng},\ and\ \citenamefont {Sandvik}}]{HShao2017}%
  \BibitemOpen
  \bibfield  {author} {\bibinfo {author} {\bibfnamefont {H.}~\bibnamefont
  {Shao}}, \bibinfo {author} {\bibfnamefont {Y.~Q.}\ \bibnamefont {Qin}},
  \bibinfo {author} {\bibfnamefont {S.}~\bibnamefont {Capponi}}, \bibinfo
  {author} {\bibfnamefont {S.}~\bibnamefont {Chesi}}, \bibinfo {author}
  {\bibfnamefont {Z.~Y.}\ \bibnamefont {Meng}},\ and\ \bibinfo {author}
  {\bibfnamefont {A.~W.}\ \bibnamefont {Sandvik}},\ }\bibfield  {title}
  {\bibinfo {title} {Nearly deconfined spinon excitations in the square-lattice
  spin-$1/2$ heisenberg antiferromagnet},\ }\href
  {https://doi.org/10.1103/PhysRevX.7.041072} {\bibfield  {journal} {\bibinfo
  {journal} {Phys. Rev. X}\ }\textbf {\bibinfo {volume} {7}},\ \bibinfo {pages}
  {041072} (\bibinfo {year} {2017})}\BibitemShut {NoStop}%
\bibitem [{\citenamefont {Sun}\ \emph {et~al.}(2018)\citenamefont {Sun},
  \citenamefont {Wang}, \citenamefont {Fang}, \citenamefont {Qi}, \citenamefont
  {Cheng},\ and\ \citenamefont {Meng}}]{GYSun2018}%
  \BibitemOpen
  \bibfield  {author} {\bibinfo {author} {\bibfnamefont {G.-Y.}\ \bibnamefont
  {Sun}}, \bibinfo {author} {\bibfnamefont {Y.-C.}\ \bibnamefont {Wang}},
  \bibinfo {author} {\bibfnamefont {C.}~\bibnamefont {Fang}}, \bibinfo {author}
  {\bibfnamefont {Y.}~\bibnamefont {Qi}}, \bibinfo {author} {\bibfnamefont
  {M.}~\bibnamefont {Cheng}},\ and\ \bibinfo {author} {\bibfnamefont {Z.~Y.}\
  \bibnamefont {Meng}},\ }\bibfield  {title} {\bibinfo {title} {Dynamical
  signature of symmetry fractionalization in frustrated magnets},\ }\href
  {https://doi.org/10.1103/PhysRevLett.121.077201} {\bibfield  {journal}
  {\bibinfo  {journal} {Phys. Rev. Lett.}\ }\textbf {\bibinfo {volume} {121}},\
  \bibinfo {pages} {077201} (\bibinfo {year} {2018})}\BibitemShut {NoStop}%
\bibitem [{\citenamefont {Ma}\ \emph {et~al.}(2018)\citenamefont {Ma},
  \citenamefont {Sun}, \citenamefont {You}, \citenamefont {Xu}, \citenamefont
  {Vishwanath}, \citenamefont {Sandvik},\ and\ \citenamefont
  {Meng}}]{NSMa2018}%
  \BibitemOpen
  \bibfield  {author} {\bibinfo {author} {\bibfnamefont {N.}~\bibnamefont
  {Ma}}, \bibinfo {author} {\bibfnamefont {G.-Y.}\ \bibnamefont {Sun}},
  \bibinfo {author} {\bibfnamefont {Y.-Z.}\ \bibnamefont {You}}, \bibinfo
  {author} {\bibfnamefont {C.}~\bibnamefont {Xu}}, \bibinfo {author}
  {\bibfnamefont {A.}~\bibnamefont {Vishwanath}}, \bibinfo {author}
  {\bibfnamefont {A.~W.}\ \bibnamefont {Sandvik}},\ and\ \bibinfo {author}
  {\bibfnamefont {Z.~Y.}\ \bibnamefont {Meng}},\ }\bibfield  {title} {\bibinfo
  {title} {Dynamical signature of fractionalization at a deconfined quantum
  critical point},\ }\href {https://doi.org/10.1103/PhysRevB.98.174421}
  {\bibfield  {journal} {\bibinfo  {journal} {Phys. Rev. B}\ }\textbf {\bibinfo
  {volume} {98}},\ \bibinfo {pages} {174421} (\bibinfo {year}
  {2018})}\BibitemShut {NoStop}%
\bibitem [{\citenamefont {Zhou}\ \emph {et~al.}(2021)\citenamefont {Zhou},
  \citenamefont {Yan}, \citenamefont {Wu}, \citenamefont {Sun}, \citenamefont
  {Starykh},\ and\ \citenamefont {Meng}}]{zhou2020amplitude}%
  \BibitemOpen
  \bibfield  {author} {\bibinfo {author} {\bibfnamefont {C.}~\bibnamefont
  {Zhou}}, \bibinfo {author} {\bibfnamefont {Z.}~\bibnamefont {Yan}}, \bibinfo
  {author} {\bibfnamefont {H.-Q.}\ \bibnamefont {Wu}}, \bibinfo {author}
  {\bibfnamefont {K.}~\bibnamefont {Sun}}, \bibinfo {author} {\bibfnamefont
  {O.~A.}\ \bibnamefont {Starykh}},\ and\ \bibinfo {author} {\bibfnamefont
  {Z.~Y.}\ \bibnamefont {Meng}},\ }\bibfield  {title} {\bibinfo {title}
  {Amplitude mode in quantum magnets via dimensional crossover},\ }\href
  {https://doi.org/10.1103/PhysRevLett.126.227201} {\bibfield  {journal}
  {\bibinfo  {journal} {Phys. Rev. Lett.}\ }\textbf {\bibinfo {volume} {126}},\
  \bibinfo {pages} {227201} (\bibinfo {year} {2021})}\BibitemShut {NoStop}%
\bibitem [{\citenamefont {Yan}\ \emph {et~al.}(2021)\citenamefont {Yan},
  \citenamefont {Wang}, \citenamefont {Ma}, \citenamefont {Qi},\ and\
  \citenamefont {Meng}}]{ZYan2021}%
  \BibitemOpen
  \bibfield  {author} {\bibinfo {author} {\bibfnamefont {Z.}~\bibnamefont
  {Yan}}, \bibinfo {author} {\bibfnamefont {Y.-C.}\ \bibnamefont {Wang}},
  \bibinfo {author} {\bibfnamefont {N.}~\bibnamefont {Ma}}, \bibinfo {author}
  {\bibfnamefont {Y.}~\bibnamefont {Qi}},\ and\ \bibinfo {author}
  {\bibfnamefont {Z.~Y.}\ \bibnamefont {Meng}},\ }\bibfield  {title} {\bibinfo
  {title} {Topological phase transition and single/multi anyon dynamics of z2
  spin liquid},\ }\href {https://doi.org/10.1038/s41535-021-00338-1} {\bibfield
   {journal} {\bibinfo  {journal} {npj Quantum Materials}\ }\textbf {\bibinfo
  {volume} {6}},\ \bibinfo {pages} {39} (\bibinfo {year} {2021})}\BibitemShut
  {NoStop}%
\bibitem [{\citenamefont {Hu}\ \emph {et~al.}(2020)\citenamefont {Hu},
  \citenamefont {Ma}, \citenamefont {Liao}, \citenamefont {Li}, \citenamefont
  {Ma}, \citenamefont {Cui}, \citenamefont {Shangguan}, \citenamefont {Huang},
  \citenamefont {Qi}, \citenamefont {Li} \emph {et~al.}}]{hu2020evidence}%
  \BibitemOpen
  \bibfield  {author} {\bibinfo {author} {\bibfnamefont {Z.}~\bibnamefont
  {Hu}}, \bibinfo {author} {\bibfnamefont {Z.}~\bibnamefont {Ma}}, \bibinfo
  {author} {\bibfnamefont {Y.-D.}\ \bibnamefont {Liao}}, \bibinfo {author}
  {\bibfnamefont {H.}~\bibnamefont {Li}}, \bibinfo {author} {\bibfnamefont
  {C.}~\bibnamefont {Ma}}, \bibinfo {author} {\bibfnamefont {Y.}~\bibnamefont
  {Cui}}, \bibinfo {author} {\bibfnamefont {Y.}~\bibnamefont {Shangguan}},
  \bibinfo {author} {\bibfnamefont {Z.}~\bibnamefont {Huang}}, \bibinfo
  {author} {\bibfnamefont {Y.}~\bibnamefont {Qi}}, \bibinfo {author}
  {\bibfnamefont {W.}~\bibnamefont {Li}}, \emph {et~al.},\ }\bibfield  {title}
  {\bibinfo {title} {Evidence of the berezinskii-kosterlitz-thouless phase in a
  frustrated magnet},\ }\href
  {https://www.nature.com/articles/s41467-020-19380-x} {\bibfield  {journal}
  {\bibinfo  {journal} {Nature communications}\ }\textbf {\bibinfo {volume}
  {11}},\ \bibinfo {pages} {1} (\bibinfo {year} {2020})}\BibitemShut {NoStop}%
\bibitem [{\citenamefont {Saito}\ \emph {et~al.}(2021)\citenamefont {Saito},
  \citenamefont {Yang}, \citenamefont {Ge}, \citenamefont {Liu}, \citenamefont
  {Taniguchi}, \citenamefont {Watanabe}, \citenamefont {Li}, \citenamefont
  {Berg},\ and\ \citenamefont {Young}}]{Saito2021}%
  \BibitemOpen
  \bibfield  {author} {\bibinfo {author} {\bibfnamefont {Y.}~\bibnamefont
  {Saito}}, \bibinfo {author} {\bibfnamefont {F.}~\bibnamefont {Yang}},
  \bibinfo {author} {\bibfnamefont {J.}~\bibnamefont {Ge}}, \bibinfo {author}
  {\bibfnamefont {X.}~\bibnamefont {Liu}}, \bibinfo {author} {\bibfnamefont
  {T.}~\bibnamefont {Taniguchi}}, \bibinfo {author} {\bibfnamefont
  {K.}~\bibnamefont {Watanabe}}, \bibinfo {author} {\bibfnamefont {J.~I.~A.}\
  \bibnamefont {Li}}, \bibinfo {author} {\bibfnamefont {E.}~\bibnamefont
  {Berg}},\ and\ \bibinfo {author} {\bibfnamefont {A.~F.}\ \bibnamefont
  {Young}},\ }\bibfield  {title} {\bibinfo {title} {Isospin pomeranchuk effect
  in twisted bilayer graphene},\ }\href
  {https://doi.org/10.1038/s41586-021-03409-2} {\bibfield  {journal} {\bibinfo
  {journal} {Nature}\ }\textbf {\bibinfo {volume} {592}},\ \bibinfo {pages}
  {220 } (\bibinfo {year} {2021})}\BibitemShut {NoStop}%
\bibitem [{\citenamefont {Bernevig}\ \emph
  {et~al.}(2021{\natexlab{b}})\citenamefont {Bernevig}, \citenamefont {Song},
  \citenamefont {Regnault},\ and\ \citenamefont {Lian}}]{bernevig2020tbg1}%
  \BibitemOpen
  \bibfield  {author} {\bibinfo {author} {\bibfnamefont {B.~A.}\ \bibnamefont
  {Bernevig}}, \bibinfo {author} {\bibfnamefont {Z.-D.}\ \bibnamefont {Song}},
  \bibinfo {author} {\bibfnamefont {N.}~\bibnamefont {Regnault}},\ and\
  \bibinfo {author} {\bibfnamefont {B.}~\bibnamefont {Lian}},\ }\bibfield
  {title} {\bibinfo {title} {Twisted bilayer graphene. i. matrix elements,
  approximations, perturbation theory, and a
  $k\ifmmode\cdot\else\textperiodcentered\fi{}p$ two-band model},\ }\href
  {https://doi.org/10.1103/PhysRevB.103.205411} {\bibfield  {journal} {\bibinfo
   {journal} {Phys. Rev. B}\ }\textbf {\bibinfo {volume} {103}},\ \bibinfo
  {pages} {205411} (\bibinfo {year} {2021}{\natexlab{b}})}\BibitemShut
  {NoStop}%
\bibitem [{\citenamefont {Song}\ \emph {et~al.}(2021)\citenamefont {Song},
  \citenamefont {Lian}, \citenamefont {Regnault},\ and\ \citenamefont
  {Bernevig}}]{song2020tbg2}%
  \BibitemOpen
  \bibfield  {author} {\bibinfo {author} {\bibfnamefont {Z.-D.}\ \bibnamefont
  {Song}}, \bibinfo {author} {\bibfnamefont {B.}~\bibnamefont {Lian}}, \bibinfo
  {author} {\bibfnamefont {N.}~\bibnamefont {Regnault}},\ and\ \bibinfo
  {author} {\bibfnamefont {B.~A.}\ \bibnamefont {Bernevig}},\ }\bibfield
  {title} {\bibinfo {title} {Twisted bilayer graphene. ii. stable symmetry
  anomaly},\ }\href {https://doi.org/10.1103/PhysRevB.103.205412} {\bibfield
  {journal} {\bibinfo  {journal} {Phys. Rev. B}\ }\textbf {\bibinfo {volume}
  {103}},\ \bibinfo {pages} {205412} (\bibinfo {year} {2021})}\BibitemShut
  {NoStop}%
\bibitem [{\citenamefont {Bernevig}\ \emph
  {et~al.}(2021{\natexlab{c}})\citenamefont {Bernevig}, \citenamefont {Song},
  \citenamefont {Regnault},\ and\ \citenamefont {Lian}}]{bernevig2020tbg3}%
  \BibitemOpen
  \bibfield  {author} {\bibinfo {author} {\bibfnamefont {B.~A.}\ \bibnamefont
  {Bernevig}}, \bibinfo {author} {\bibfnamefont {Z.-D.}\ \bibnamefont {Song}},
  \bibinfo {author} {\bibfnamefont {N.}~\bibnamefont {Regnault}},\ and\
  \bibinfo {author} {\bibfnamefont {B.}~\bibnamefont {Lian}},\ }\bibfield
  {title} {\bibinfo {title} {Twisted bilayer graphene. iii. interacting
  hamiltonian and exact symmetries},\ }\href
  {https://doi.org/10.1103/PhysRevB.103.205413} {\bibfield  {journal} {\bibinfo
   {journal} {Phys. Rev. B}\ }\textbf {\bibinfo {volume} {103}},\ \bibinfo
  {pages} {205413} (\bibinfo {year} {2021}{\natexlab{c}})}\BibitemShut
  {NoStop}%
\bibitem [{\citenamefont {Tarnopolsky}\ \emph {et~al.}(2019)\citenamefont
  {Tarnopolsky}, \citenamefont {Kruchkov},\ and\ \citenamefont
  {Vishwanath}}]{Ashvin2019}%
  \BibitemOpen
  \bibfield  {author} {\bibinfo {author} {\bibfnamefont {G.}~\bibnamefont
  {Tarnopolsky}}, \bibinfo {author} {\bibfnamefont {A.~J.}\ \bibnamefont
  {Kruchkov}},\ and\ \bibinfo {author} {\bibfnamefont {A.}~\bibnamefont
  {Vishwanath}},\ }\bibfield  {title} {\bibinfo {title} {Origin of magic angles
  in twisted bilayer graphene},\ }\href
  {https://doi.org/10.1103/PhysRevLett.122.106405} {\bibfield  {journal}
  {\bibinfo  {journal} {Phys. Rev. Lett.}\ }\textbf {\bibinfo {volume} {122}},\
  \bibinfo {pages} {106405} (\bibinfo {year} {2019})}\BibitemShut {NoStop}%
\bibitem [{\citenamefont {Assaad}\ and\ \citenamefont
  {Evertz}(2008)}]{Assaad2008}%
  \BibitemOpen
  \bibfield  {author} {\bibinfo {author} {\bibfnamefont {F.}~\bibnamefont
  {Assaad}}\ and\ \bibinfo {author} {\bibfnamefont {H.}~\bibnamefont
  {Evertz}},\ }\bibinfo {title} {World-line and determinantal quantum monte
  carlo methods for spins, phonons and electrons},\ in\ \href
  {https://doi.org/10.1007/978-3-540-74686-7_10} {\emph {\bibinfo {booktitle}
  {Computational Many-Particle Physics}}},\ \bibinfo {editor} {edited by\
  \bibinfo {editor} {\bibfnamefont {H.}~\bibnamefont {Fehske}}, \bibinfo
  {editor} {\bibfnamefont {R.}~\bibnamefont {Schneider}},\ and\ \bibinfo
  {editor} {\bibfnamefont {A.}~\bibnamefont {Wei{\ss}e}}}\ (\bibinfo
  {publisher} {Springer Berlin Heidelberg},\ \bibinfo {address} {Berlin,
  Heidelberg},\ \bibinfo {year} {2008})\ pp.\ \bibinfo {pages}
  {277--356}\BibitemShut {NoStop}%
\bibitem [{\citenamefont {Da~Liao}\ \emph {et~al.}(2019)\citenamefont
  {Da~Liao}, \citenamefont {Meng},\ and\ \citenamefont {Xu}}]{YDLiao2019PRL}%
  \BibitemOpen
  \bibfield  {author} {\bibinfo {author} {\bibfnamefont {Y.}~\bibnamefont
  {Da~Liao}}, \bibinfo {author} {\bibfnamefont {Z.~Y.}\ \bibnamefont {Meng}},\
  and\ \bibinfo {author} {\bibfnamefont {X.~Y.}\ \bibnamefont {Xu}},\
  }\bibfield  {title} {\bibinfo {title} {Valence bond orders at charge
  neutrality in a possible two-orbital extended hubbard model for twisted
  bilayer graphene},\ }\href {https://doi.org/10.1103/PhysRevLett.123.157601}
  {\bibfield  {journal} {\bibinfo  {journal} {Phys. Rev. Lett.}\ }\textbf
  {\bibinfo {volume} {123}},\ \bibinfo {pages} {157601} (\bibinfo {year}
  {2019})}\BibitemShut {NoStop}%
\bibitem [{sup()}]{suppl}%
  \BibitemOpen
  \href@noop {} {\bibinfo  {journal} {The momentum space QMC methodology, the
  implementation of order measurements within the QMC, brief description of the
  stochastic analytic continuation and exact many-body excitations at chiral
  limits, are presented in this Supplemental Material}\ }\BibitemShut {NoStop}%
\bibitem [{\citenamefont {{Vafek}}\ and\ \citenamefont
  {{Kang}}(2021)}]{Vafek2021}%
  \BibitemOpen
\bibfield  {journal} {  }\bibfield  {author} {\bibinfo {author} {\bibfnamefont
  {O.}~\bibnamefont {{Vafek}}}\ and\ \bibinfo {author} {\bibfnamefont
  {J.}~\bibnamefont {{Kang}}},\ }\bibfield  {title} {\bibinfo {title} {{Lattice
  model for the Coulomb interacting chiral limit of the magic angle twisted
  bilayer graphene: symmetries, obstructions and excitations}},\ }\href@noop {}
  {\bibfield  {journal} {\bibinfo  {journal} {arXiv e-prints}\ ,\ \bibinfo
  {eid} {arXiv:2106.05670}} (\bibinfo {year} {2021})},\ \Eprint
  {https://arxiv.org/abs/2106.05670} {arXiv:2106.05670 [cond-mat.str-el]}
  \BibitemShut {NoStop}%
\bibitem [{\citenamefont {Sandvik}(1998)}]{Sandvik1998}%
  \BibitemOpen
  \bibfield  {author} {\bibinfo {author} {\bibfnamefont {A.~W.}\ \bibnamefont
  {Sandvik}},\ }\bibfield  {title} {\bibinfo {title} {Stochastic method for
  analytic continuation of quantum monte carlo data},\ }\href
  {https://doi.org/10.1103/PhysRevB.57.10287} {\bibfield  {journal} {\bibinfo
  {journal} {Phys. Rev. B}\ }\textbf {\bibinfo {volume} {57}},\ \bibinfo
  {pages} {10287} (\bibinfo {year} {1998})}\BibitemShut {NoStop}%
\bibitem [{\citenamefont {Beach}(2004)}]{beach2004identifying}%
  \BibitemOpen
  \bibfield  {author} {\bibinfo {author} {\bibfnamefont {K.}~\bibnamefont
  {Beach}},\ }\bibfield  {title} {\bibinfo {title} {Identifying the maximum
  entropy method as a special limit of stochastic analytic continuation},\
  }\href {https://arxiv.org/abs/cond-mat/0403055} {\bibfield  {journal}
  {\bibinfo  {journal} {arXiv preprint cond-mat/0403055}\ } (\bibinfo {year}
  {2004})}\BibitemShut {NoStop}%
\bibitem [{\citenamefont {Sylju\aa{}sen}(2008)}]{Olav2008}%
  \BibitemOpen
  \bibfield  {author} {\bibinfo {author} {\bibfnamefont {O.~F.}\ \bibnamefont
  {Sylju\aa{}sen}},\ }\bibfield  {title} {\bibinfo {title} {Using the average
  spectrum method to extract dynamics from quantum monte carlo simulations},\
  }\href {https://doi.org/10.1103/PhysRevB.78.174429} {\bibfield  {journal}
  {\bibinfo  {journal} {Phys. Rev. B}\ }\textbf {\bibinfo {volume} {78}},\
  \bibinfo {pages} {174429} (\bibinfo {year} {2008})}\BibitemShut {NoStop}%
\bibitem [{\citenamefont {Li}\ \emph {et~al.}(2020)\citenamefont {Li},
  \citenamefont {Da~Liao}, \citenamefont {Chen}, \citenamefont {Zeng},
  \citenamefont {Sheng}, \citenamefont {Qi}, \citenamefont {Meng},\ and\
  \citenamefont {Li}}]{li2020kosterlitz}%
  \BibitemOpen
  \bibfield  {author} {\bibinfo {author} {\bibfnamefont {H.}~\bibnamefont
  {Li}}, \bibinfo {author} {\bibfnamefont {Y.}~\bibnamefont {Da~Liao}},
  \bibinfo {author} {\bibfnamefont {B.-B.}\ \bibnamefont {Chen}}, \bibinfo
  {author} {\bibfnamefont {X.-T.}\ \bibnamefont {Zeng}}, \bibinfo {author}
  {\bibfnamefont {X.-L.}\ \bibnamefont {Sheng}}, \bibinfo {author}
  {\bibfnamefont {Y.}~\bibnamefont {Qi}}, \bibinfo {author} {\bibfnamefont
  {Z.~Y.}\ \bibnamefont {Meng}},\ and\ \bibinfo {author} {\bibfnamefont
  {W.}~\bibnamefont {Li}},\ }\bibfield  {title} {\bibinfo {title}
  {Kosterlitz-thouless melting of magnetic order in the triangular quantum
  ising material tmmggao 4},\ }\href
  {https://www.nature.com/articles/s41467-020-14907-8} {\bibfield  {journal}
  {\bibinfo  {journal} {Nature communications}\ }\textbf {\bibinfo {volume}
  {11}},\ \bibinfo {pages} {1} (\bibinfo {year} {2020})}\BibitemShut {NoStop}%
\bibitem [{\citenamefont {Jiang}\ \emph {et~al.}(2021)\citenamefont {Jiang},
  \citenamefont {Liu}, \citenamefont {Klein}, \citenamefont {Wang},
  \citenamefont {Sun}, \citenamefont {Chubukov},\ and\ \citenamefont
  {Meng}}]{jiang2020}%
  \BibitemOpen
  \bibfield  {author} {\bibinfo {author} {\bibfnamefont {W.}~\bibnamefont
  {Jiang}}, \bibinfo {author} {\bibfnamefont {Y.}~\bibnamefont {Liu}}, \bibinfo
  {author} {\bibfnamefont {A.}~\bibnamefont {Klein}}, \bibinfo {author}
  {\bibfnamefont {Y.}~\bibnamefont {Wang}}, \bibinfo {author} {\bibfnamefont
  {K.}~\bibnamefont {Sun}}, \bibinfo {author} {\bibfnamefont {A.~V.}\
  \bibnamefont {Chubukov}},\ and\ \bibinfo {author} {\bibfnamefont {Z.~Y.}\
  \bibnamefont {Meng}},\ }\bibfield  {title} {\bibinfo {title} {Pseudogap and
  superconductivity emerging from quantum magnetic fluctuations: a monte carlo
  study},\ }\href {https://arxiv.org/abs/2105.03639} {\bibfield  {journal}
  {\bibinfo  {journal} {arXiv preprint arXiv:2105.03639}\ } (\bibinfo {year}
  {2021})}\BibitemShut {NoStop}%
\bibitem [{\citenamefont {Chen}\ \emph
  {et~al.}(2021{\natexlab{b}})\citenamefont {Chen}, \citenamefont {Yuan},
  \citenamefont {Qi},\ and\ \citenamefont {Meng}}]{ChuangChen2021}%
  \BibitemOpen
  \bibfield  {author} {\bibinfo {author} {\bibfnamefont {C.}~\bibnamefont
  {Chen}}, \bibinfo {author} {\bibfnamefont {T.}~\bibnamefont {Yuan}}, \bibinfo
  {author} {\bibfnamefont {Y.}~\bibnamefont {Qi}},\ and\ \bibinfo {author}
  {\bibfnamefont {Z.~Y.}\ \bibnamefont {Meng}},\ }\bibfield  {title} {\bibinfo
  {title} {Fermi arcs and pseudogap in a lattice model of a doped orthogonal
  metal},\ }\href {https://doi.org/10.1103/PhysRevB.103.165131} {\bibfield
  {journal} {\bibinfo  {journal} {Phys. Rev. B}\ }\textbf {\bibinfo {volume}
  {103}},\ \bibinfo {pages} {165131} (\bibinfo {year}
  {2021}{\natexlab{b}})}\BibitemShut {NoStop}%
\bibitem [{\citenamefont {Nam}\ and\ \citenamefont {Koshino}(2017)}]{Nam2017}%
  \BibitemOpen
  \bibfield  {author} {\bibinfo {author} {\bibfnamefont {N.~N.~T.}\
  \bibnamefont {Nam}}\ and\ \bibinfo {author} {\bibfnamefont {M.}~\bibnamefont
  {Koshino}},\ }\bibfield  {title} {\bibinfo {title} {Lattice relaxation and
  energy band modulation in twisted bilayer graphene},\ }\href
  {https://doi.org/10.1103/PhysRevB.96.075311} {\bibfield  {journal} {\bibinfo
  {journal} {Phys. Rev. B}\ }\textbf {\bibinfo {volume} {96}},\ \bibinfo
  {pages} {075311} (\bibinfo {year} {2017})}\BibitemShut {NoStop}%
\bibitem [{\citenamefont {Carr}\ \emph {et~al.}(2019)\citenamefont {Carr},
  \citenamefont {Fang}, \citenamefont {Po}, \citenamefont {Vishwanath},\ and\
  \citenamefont {Kaxiras}}]{Ashvin20192}%
  \BibitemOpen
  \bibfield  {author} {\bibinfo {author} {\bibfnamefont {S.}~\bibnamefont
  {Carr}}, \bibinfo {author} {\bibfnamefont {S.}~\bibnamefont {Fang}}, \bibinfo
  {author} {\bibfnamefont {H.~C.}\ \bibnamefont {Po}}, \bibinfo {author}
  {\bibfnamefont {A.}~\bibnamefont {Vishwanath}},\ and\ \bibinfo {author}
  {\bibfnamefont {E.}~\bibnamefont {Kaxiras}},\ }\bibfield  {title} {\bibinfo
  {title} {Derivation of wannier orbitals and minimal-basis tight-binding
  hamiltonians for twisted bilayer graphene: First-principles approach},\
  }\href {https://doi.org/10.1103/PhysRevResearch.1.033072} {\bibfield
  {journal} {\bibinfo  {journal} {Phys. Rev. Research}\ }\textbf {\bibinfo
  {volume} {1}},\ \bibinfo {pages} {033072} (\bibinfo {year}
  {2019})}\BibitemShut {NoStop}%
\bibitem [{\citenamefont {Feldner}\ \emph {et~al.}(2011)\citenamefont
  {Feldner}, \citenamefont {Meng}, \citenamefont {Lang}, \citenamefont
  {Assaad}, \citenamefont {Wessel},\ and\ \citenamefont
  {Honecker}}]{Feldner2011}%
  \BibitemOpen
  \bibfield  {author} {\bibinfo {author} {\bibfnamefont {H.}~\bibnamefont
  {Feldner}}, \bibinfo {author} {\bibfnamefont {Z.~Y.}\ \bibnamefont {Meng}},
  \bibinfo {author} {\bibfnamefont {T.~C.}\ \bibnamefont {Lang}}, \bibinfo
  {author} {\bibfnamefont {F.~F.}\ \bibnamefont {Assaad}}, \bibinfo {author}
  {\bibfnamefont {S.}~\bibnamefont {Wessel}},\ and\ \bibinfo {author}
  {\bibfnamefont {A.}~\bibnamefont {Honecker}},\ }\bibfield  {title} {\bibinfo
  {title} {Dynamical signatures of edge-state magnetism on graphene
  nanoribbons},\ }\href {https://doi.org/10.1103/PhysRevLett.106.226401}
  {\bibfield  {journal} {\bibinfo  {journal} {Phys. Rev. Lett.}\ }\textbf
  {\bibinfo {volume} {106}},\ \bibinfo {pages} {226401} (\bibinfo {year}
  {2011})}\BibitemShut {NoStop}%
\bibitem [{\citenamefont {Golor}\ \emph {et~al.}(2013)\citenamefont {Golor},
  \citenamefont {Lang},\ and\ \citenamefont {Wessel}}]{Golor2013}%
  \BibitemOpen
  \bibfield  {author} {\bibinfo {author} {\bibfnamefont {M.}~\bibnamefont
  {Golor}}, \bibinfo {author} {\bibfnamefont {T.~C.}\ \bibnamefont {Lang}},\
  and\ \bibinfo {author} {\bibfnamefont {S.}~\bibnamefont {Wessel}},\
  }\bibfield  {title} {\bibinfo {title} {Quantum monte carlo studies of edge
  magnetism in chiral graphene nanoribbons},\ }\href
  {https://doi.org/10.1103/PhysRevB.87.155441} {\bibfield  {journal} {\bibinfo
  {journal} {Phys. Rev. B}\ }\textbf {\bibinfo {volume} {87}},\ \bibinfo
  {pages} {155441} (\bibinfo {year} {2013})}\BibitemShut {NoStop}%
\bibitem [{\citenamefont {Golor}\ \emph {et~al.}(2014)\citenamefont {Golor},
  \citenamefont {Wessel},\ and\ \citenamefont {Schmidt}}]{Golor2014}%
  \BibitemOpen
  \bibfield  {author} {\bibinfo {author} {\bibfnamefont {M.}~\bibnamefont
  {Golor}}, \bibinfo {author} {\bibfnamefont {S.}~\bibnamefont {Wessel}},\ and\
  \bibinfo {author} {\bibfnamefont {M.~J.}\ \bibnamefont {Schmidt}},\
  }\bibfield  {title} {\bibinfo {title} {Quantum nature of edge magnetism in
  graphene},\ }\href {https://doi.org/10.1103/PhysRevLett.112.046601}
  {\bibfield  {journal} {\bibinfo  {journal} {Phys. Rev. Lett.}\ }\textbf
  {\bibinfo {volume} {112}},\ \bibinfo {pages} {046601} (\bibinfo {year}
  {2014})}\BibitemShut {NoStop}%
\end{thebibliography}%
	
\newpage
\begin{appendix}
	\begin{widetext}
\section{Supplemental Material for  \\[0.5em] Dynamical properties of collective excitations in twisted bilayer Graphene }

\subsection{Section I: Momentum space QMC methodology}
Following the description in Ref.~\cite{XuZhang2021}, in this section, we elucidate the momentum space quantum Monte Carlo method in detail. 

First, the partition function of the TBG Hamiltonian in Eq. (3) of the main text is given by:
\begin{equation}
	\begin{aligned}
		Z&=\operatorname{Tr}\left[e^{-\beta H}\right]\\ &=\operatorname{Tr}\left[\left(e^{-\Delta \tau H} \right)^{L_{\tau}}\right]\\
		&=\operatorname{Tr}\left[\prod_{\tau=1}^{L_\tau}e^{-\Delta \tau H_0}e^{-\Delta\tau H_{int}}\right]+O(\Delta \tau^2)
	\end{aligned}
\end{equation}
For the interaction part $H_{i n t}=\frac{1}{2 \Omega} \sum_{\mathbf{q}, \mathbf{G},|\mathbf{q}+\mathbf{G}| \neq 0} V(\mathbf{q}+\mathbf{G}) \delta \rho_{\mathbf{q}+\mathbf{G}} \delta \rho_{-\mathbf{q}-\mathbf{G}}$, we have

\begin{equation}
	\sum_{\mathbf{q}, \mathbf{G},|\mathbf{q}+\mathbf{G}| \neq 0}\frac{1}{2\Omega} V(\mathbf{q}+\mathbf{G}) \delta \rho_{\mathbf{q}+\mathbf{G}} \delta \rho_{-\mathbf{q}-\mathbf{G}}=\sum_{|\mathbf{q}+\mathbf{G}| \neq 0} \frac{V(\mathbf{q}+\mathbf{G})}{4\Omega}\left[\left(\delta \rho_{-\mathbf{q}-\mathbf{G}}+\delta \rho_{\mathbf{q}+\mathbf{G}}\right)^{2}-\left(\delta \rho_{-\mathbf{q}-\mathbf{G}}-\delta \rho_{\mathbf{q}+\mathbf{G}}\right)^{2}\right]
\end{equation}
then
\begin{equation}
	\begin{aligned}
		e^{-\Delta \tau \hat{H}_{int}}&= \prod_{|\bq+\bG| \neq 0}  e^{-\Delta \frac{V(\mathbf{q}+\mathbf{G})}{4\Omega}\left[\left(\delta \rho_{-\mathbf{q}-\mathbf{G}}+\delta \rho_{\mathbf{q}+\mathbf{G}}\right)^{2}-\left(\delta \rho_{-\mathbf{q}-\mathbf{G}}-\delta \rho_{\mathbf{q}+\mathbf{G}}\right)^{2}\right]}.
	\end{aligned}
\end{equation}

The discrete Hubbard-Stratonovich transformation~\cite{Assaad2008,YDLiao2021PRX,YDLiao2019PRL,XuZhang2021} reads:
\begin{equation}
	e^{\alpha \hat{O}^{2}}=\frac{1}{4} \sum_{l=\pm 1,\pm 2} \gamma(l) e^{\sqrt{\alpha} \eta(l) \hat{o}}+O\left(\alpha^{4}\right)  
	\label{eq:eq8}
\end{equation}
where $l=\pm 1,\pm 2$, 
and \begin{equation}\begin{aligned}
		\gamma(\pm 1)=1+\sqrt{6} / 3, & \qquad \gamma(\pm 2)=1-\sqrt{6} / 3 \\
		\eta(\pm 1)=\pm \sqrt{2(3-\sqrt{6})}, &\qquad \eta(\pm 2)=\pm \sqrt{2(3+\sqrt{6})}
	\end{aligned}
\end{equation}

This can be seen from the following simple derivation. Assuming, \begin{equation}
	\gamma(1)=\gamma(-1)=a, \quad \gamma(2)=\gamma(-2)=b, \quad \eta(1)=\sqrt{c}=-\eta(1), \quad \eta(2)=\sqrt{d}=-\eta(2)
\end{equation}

Taylor expands both sides of Eq.~\eqref{eq:eq8} to $O\left(\alpha^{4}\right)$ and compare the coefficients, we obtain:
\begin{equation}
	1=\frac{1}{2}(a+b),\quad
	1=\frac{1}{4}(a c+b d) ,\quad
	\frac{1}{2}=\frac{1}{48}\left(a c^{2}+b d^{2}\right),\quad
	\frac{1}{6}=\frac{1}{1440}\left(a c^{3}+b d^{3}\right)
\end{equation}
solve these equations, then we have:
\begin{equation}
	\begin{aligned}
		&a=1+\sqrt{6} / 3, \quad b=1-\sqrt{6} / 3 \\
		&c=2(3-\sqrt{6}), \quad d=2(3+\sqrt{6})
	\end{aligned}
\end{equation}
as those in Eq.~\eqref{eq:eq8}. 

For a fermion bilinear, i.e. free fermion system, its partition function can be expressed as a determinant,
\begin{equation}
	\operatorname{Tr}\left[e^{-\sum_{i, j} c_{i}^{\dagger} A_{i, j c} c_{j}-\sum_{i, j} c_{i}^{\dagger} B_{i, j} c_{j}}\right]=\operatorname{Det}\left(1+e^{-\mathbf{A}} e^{-\mathbf{B}}\right). 
	\label{eq:eq13}
\end{equation}	
Put Eqs.~\eqref{eq:eq8} and ~\eqref{eq:eq13} together, the partition function of our interacting TBG system can be expressed as:	

\begin{equation}
	\begin{aligned}			
		Z &=\sum_{\left\{ l_{|\bq+\bG|,a,\tau}=\pm 1,\pm 2\right\} } \prod_{\tau=1}^{L_\tau} e^{-\Delta \tau H_0} \operatorname{Tr}_{c}\left[\prod_{|\bq+\bG|\neq 0}  \frac{1}{16} \gamma\left(l_{|\bq+\bG|,1,\tau}\right)\gamma\left(l_{|\bq+\bG|,2,\tau}\right) e^{i \eta\left(l_{|\mathbf{q}|_{1}, t}\right) A_{\mathbf{q}}\left(\delta \rho_{-\mathbf{q}}+\delta \rho_{\mathbf{q}}\right)} e^{\eta\left(l_{|\mathbf{q}|_{2}, t}\right) A_{\mathbf{q}}\left(\delta \rho_{-\mathbf{q}}-\delta \rho_{\mathbf{q}}\right)} \right] \\
		&\qquad \qquad + O(\Delta \tau {}^2)
	\end{aligned}
\end{equation}
where $A_{\mathbf{q}+\mathbf{G}}=\sqrt{\frac{\Delta \tau}{4} \frac{V(\mathbf{q}+\mathbf{G})}{\Omega}}$ and the trace over fermion operators gives rise to the determinant for each auxiliary configuration. The free of sign-problem and the Monte Carlo sampling scheme are presented in Ref.~\cite{XuZhang2021}.

\subsection{Section II: Order Parameter}
As discussed in the main text. For the correlation functions of VP order parameter, we define 
\begin{equation}
	\begin{aligned}
		S_{VP}(\boldsymbol{q}) & \equiv \frac{1}{N^{2}}\left\langle\mathcal{O}_{a}(-\boldsymbol{q}) \mathcal{O}_{a}(\boldsymbol{q})\right\rangle \\
		&\mathcal{O}_{a}(\boldsymbol{q})  \equiv \sum_{\boldsymbol{k}} d_{\boldsymbol{k}+\boldsymbol{q}}^{\dagger} \tau_z \eta_0 d_{\boldsymbol{k}}
	\end{aligned}
\end{equation}
where $\eta_0$ is for band index and $\tau_z$ is for valley index. Then its QMC implementation reads as,
\begin{equation}
	\begin{aligned}
		S_{VP}(q)=&\frac{1}{N^{2}}   \sum_{k_{1},k_{2}}\sum_{n_{1},n_{2}}\sum_{\tau_1,\tau_2=\pm} \left(\tau_1 \tau_2\right) \left \langle d_{k_{1}, n_{1}, \tau_1}^{\dagger} d_{k_{1}+q, m_{1}, \tau_1} d_{k_{2}+q, n_{2}, \tau_2}^{\dagger} d_{k_{2}, m_{2}, \tau_2} \right\rangle\\
		=&\frac{1}{N^{2}}\langle \left(\sum_{k_{1}}\left(\sum_{n_{1}}  d_{k_{1}, n_{1}, \tau}^{\dagger} d_{k_{1}+q, n_{1}, \tau}-\tilde{d}_{k_{1}, n_{1}-\tau} \tilde{d}_{k_{1}+q, n_{1},-\tau}^{\dagger}\right)\right) \\
		& \qquad \left.\cdot\left(\sum_{k_{2}}\left(\sum_{n_{2}}  d_{k_{2}+q, n_{2}, \tau}^{\dagger} d_{k_{2}, n_{2}, \tau}- \tilde{d}_{k_{2}+q, n_{2}-\tau} \tilde{d}_{k_{2}, n_{2}-\tau}^{\dagger}\right)\right)\right\rangle \\
		=&\frac{1}{N^{2}}  \sum_{k_{1},k_{2}} \sum_{n_{1},n_{2}}\mathrm{Gc}_{n_1 n_1,\tau}(k_{1},k_{1}+q)\mathrm{Gc}_{n_2 n_2,\tau}(k_{2}+q,k_{2})\\
		& \qquad \qquad + \mathrm{G}^{*}_{n_1 n_1,\tau}(k_{1},k_{1}+q)\mathrm{G}^{*}_{n_2 n_2,\tau}(k_{2}+q,k_{2})\\
		& \qquad \qquad + \mathrm{Gc}_{n_1 n_2,\tau}(k_{1},k_{2})\mathrm{G}_{n_1 n_2,\tau}(k_{1}+q,k_{2}+q)\\
		& \qquad \qquad + \mathrm{G}^{*}_{n_1 n_2,\tau}(k_{1},k_{2})\mathrm{Gc}^{*}_{n_1 n_2,\tau}(k_{1}+q,k_{2}+q)\\
		& \qquad \qquad -\mathrm{Gc}_{n_1 n_1,\tau}(k_{1},k_{1}+q)\mathrm{G}^{*}_{n_2 n_2,\tau}(k_{2}+q,k_{2})\\
		& \qquad \qquad -\mathrm{G}^{*}_{n_1 n_1,\tau}(k_{1},k_{1}+q)\mathrm{Gc}_{n_2 n_2,\tau}(k_{2}+q,k_{2})
	\end{aligned}
\end{equation}			
where $\tilde{d}_{\mathbf{k}, m,-\tau}=m * d_{\mathbf{k},-m,-\tau}^{\dagger}$ and $d^\dagger_{\mathbf{k}_1, m,-\tau} d_{\mathbf{k}_2, n,-\tau}=(mn) \tilde{d}_{\mathbf{k}_1, -m,-\tau} \tilde{d}^\dagger_{\mathbf{k}_2, -n,-\tau}=(mn)G_{-m,-n}^{*}(k_1,k_2)$, note we define the fermion Green's function as $\mathrm{G}_{ij}=\langle d^{\dagger}_i d_j \rangle$ and define $\mathrm{Gc}_{i j}=\delta_{i j}-\operatorname{G}_{j i}$. 

For the correlation function of the IVC order parameter, we define
\begin{equation}
	\begin{aligned}
		S_{IVC}(q) & \equiv \frac{1}{N^{2}}\left\langle\mathcal{O}_{a}(-\boldsymbol{q}) \mathcal{O}_{a}(\boldsymbol{q})\right\rangle \\
		&\mathcal{O}_{a}(\boldsymbol{q})  \equiv \sum_{\boldsymbol{k}} d_{\boldsymbol{k}+\boldsymbol{q}}^{\dagger} \tau_x \eta_y d_{\boldsymbol{k}}
	\end{aligned}
\end{equation}
and its QMC implementation reads as,
\begin{equation}
	\begin{aligned}
		S_{IVC}(q)=\frac{1}{N^{2}}  &  \sum_{k_{1},k_{2}}\sum_{n_{1},n_{2}, }\sum_{\tau_1,\tau_2=\pm} \left(n_1 n_2\right) \left \langle d_{k_{1}, n_{1}, \tau_1}^{\dagger} d_{k_{1}+q, -n_1, -\tau_1} d_{k_{2}+q, n_{2}, \tau_2}^{\dagger} d_{k_{2}, -n_{2}, -\tau_2} \right\rangle\\
		=\frac{1}{N^{2}} &  \sum_{k_{1},k_{2}}\sum_{n_{1},n_{2}, }\sum_{\tau=\pm} \left(n_1 n_2\right)\mathrm{Gc}_{n_1 ,-n_2,\tau}(k_{1},k_{2})\mathrm{G}_{-n_1, n_2,-\tau}(k_{1}+q,k_{2}+q)\\
		=\frac{1}{N^{2}} & \sum_{k_{1},k_{2}}\sum_{n_{1},n_{2}, } \mathrm{Gc}_{n_1,-n_2,\tau}(k_{1},k_{2})\mathrm{Gc}^{*}_{n_1, -n_2,\tau}(k_{1}+q,k_{2}+q)\\
		&\qquad \qquad + \mathrm{G}^{*}_{n_1,-n_2,\tau}(k_{1},k_{2})\mathrm{G}_{n_1, -n_2,\tau}(k_{1}+q,k_{2}+q)
	\end{aligned}
\end{equation}

\subsection{Section III: Analytic continuation}
From QMC simulations, we only obtain the imaginary time or imaginary frequency Green's functions, we further perform the stochastic analytic continuation (SAC) method~\cite{Sandvik1998,beach2004identifying,Sandvik2016,Olav2008,HShao2017,NSMa2018,zhou2020amplitude,GYSun2018,ZYan2021,hu2020evidence,li2020kosterlitz,jiang2020,XuZhang2021} to obtain the real frequency spectral function $A(k,\omega)$. 

Here we give a brief description of the scheme. 	

Firstly, we define :
$e^{-\beta \Omega}=\operatorname{Tr}\left(e^{-\beta(H-\mu N)}\right), K \equiv H-\mu N$.
The imaginary time Green's function is:
\begin{equation}\begin{aligned}
		G(\tau) &=\left\langle T_{\tau}d\(\tau\) d^\dagger\(0\) \right\rangle\\
		&=\operatorname{Tr}\left[e^{-\beta(K-\Omega)} T_{\tau} e^{\tau K} d \,e^{-\tau K} d^{\dagger}\right]
	\end{aligned}\end{equation}
	where $K|m\rangle=E_{m}|m\rangle$.
	Then if we consider the Lehmann representation:
	\begin{equation}\begin{aligned}
			\tau>0: & G(\tau)=e^{\beta \Omega} \sum_{n, m}\left\langle n\left|e^{-\beta K} d(\tau)\right| m\right\rangle\left\langle m\left|d^{\dagger}(0)\right| n\right\rangle \\
			& G(\tau)=e^{\beta \Omega} \sum_{n, m}|\langle n|d| m\rangle|^{2} e^{-\beta E_{n}} e^{\tau\left(E_{n}-E_{m}\right)}
		\end{aligned}\end{equation}
		Once again, imaginary frequency Green's function is : 
		\begin{equation}\begin{aligned}
				G\left(i \omega_{n}\right)&=\int_{0}^{\beta} d \tau e^{i \omega_{n}\tau} G(\tau) \\
				&=-e^{\beta \Omega} \sum_{n, m}|\langle n|d| m\rangle|^{2} e^{-\beta E_{n}} \frac{e^{\(i \omega_n +E_n-E_m\)\tau} |^{\beta}_{0}}{i\omega_n +E_n-E_m}\\
				&=e^{\beta \Omega} \sum_{n, m}|\langle n|d| m\rangle|^{2} \frac{e^{-\beta E_n} \mp e^{-\beta E_m}}{i\omega_n +E_n-E_m} \label{Eq:A21}
			\end{aligned}\end{equation}
			here $\mp$ for boson and fermion. And we use $e^{i \omega_n \beta}=\pm 1$.
			
			Then we carry out the analytic continuation: $i \omega_{n} \rightarrow w+i \delta$ and obtain the retarded real frequency Green's function $ G\left(i \omega_{n}\right) \rightarrow G^{ret}(\omega)$, where $G^{ret}\left( \omega \right)=\int_{-\infty}^{\infty} e^{i\omega t} G^{ret}\left(t\right) \mathrm{d}t$ and $G^{ret}\left(t-t^{\prime}\right)=-i \theta\left(t-t^{\prime}\right)\left\langle\left[d(t) d^{\dagger}\left(t^{\prime}\right)+d^{\dagger}\left(t^{\prime}\right) d(t)\right]\right\rangle$. The spectral function is obtained by the retarded Green function: $A(k,\omega)=-(1 / \pi) \operatorname{Im} G^{r e t}(k,\omega)$
			
			\begin{equation}
				\begin{aligned}
					A(k,\omega)&=-(1 / \pi) \operatorname{Im} G^{r e t}(k,\omega)\\
					&=e^{\beta \Omega} \sum_{n, m}|\langle n|d| m\rangle|^{2} \(e^{-\beta E_n} \mp e^{-\beta E_m}\)\delta(\omega +E_n-E_m) 
					\label{Eq:A22}
				\end{aligned}
			\end{equation}
			from Eqs.~\eqref{Eq:A21} and ~\eqref{Eq:A22}, we can get :
			\begin{equation}
				G(k,\tau)=\int_{-\infty}^{\infty} d \omega\left[\frac{e^{-\omega \tau}}{1\mp e^{-\beta \omega}}\right] A(k,\omega)\end{equation}
			Note again $\mp$ for boson and fermion.
			
			For boson Green function:
			\begin{equation}
				G(k,\tau)=\int_{0}^{+\infty} \mathrm{d} \omega \frac{e^{-\tau \omega}+e^{-(\beta-\tau) \omega}}{1-e^{\boldsymbol{-} \beta \omega}} A(k,\omega).
				\label{eq:Aeq24}
			\end{equation}
			
			In the spectroscopy measurements such as the inelastic neutron scattering, the spectral function $S(k, \omega)=\frac{1}{1-e^{-\beta \omega}} \operatorname{Im} \chi(k, \omega)$, where $\chi(k, \omega)$ is dynamical spin susceptibility. We can see $\operatorname{Im} \chi(k, \omega)$ is the spectral function $A(k,\omega)$ mentioned above.
			
			Now we discuss the details of stochastic analytic continuation. The idea is to give a very generic variational ansatz of the spectrum $A(k,\omega)$, and obtain corresponding Green's function $G(k,\tau)$ following Eq.~\eqref{eq:Aeq24} . Then compare the Green's function with the Green's function obtained from QMC by the quantity $\chi^2_{F/B}$. Definition of $\chi^2_{F/B}$ is
			
			\begin{equation}
				\chi_{F}^{2}=\sum_{i j}\left(\bar{G}\left(\tau_{i}\right)-\int_{-\infty}^{\infty} d \omega\left[\frac{e^{-\omega \tau_i}}{1+e^{-\beta \omega}}\right] A(\omega)\right)\left(C^{-1}\right)_{i j}\left(\bar{G}\left(\tau_{j}\right)-\int_{-\infty}^{\infty} d \omega\left[\frac{e^{-\omega \tau_j}}{1+e^{-\beta \omega}}\right] A(\omega)\right)\end{equation}
			and
			\begin{equation}
				\chi_{B}^{2}=\sum_{i j}\left(\bar{G}\left(\tau_{i}\right)-\int_{0}^{\infty} d \omega\left[\frac{e^{-\omega \tau_i}+e^{-(\beta-\tau) \omega}}{1-e^{-\beta \omega}}\right] A(\omega)\right)\left(C^{-1}\right)_{i j}\left(\bar{G}\left(\tau_{j}\right)-\int_{0}^{\infty} d \omega\left[\frac{e^{-\omega \tau_j} + e^{-(\beta-\tau) \omega} }{1-e^{-\beta \omega}}\right] A(\omega)\right)\end{equation}
			where
			\begin{equation} C_{i j}=\frac{1}{N_{b}\left(N_{b}-1\right)} \sum_{b=1}^{N_b}\left(G^{b}\left(\tau_{i}\right)-\bar{G}\left(\tau_{i}\right)\right)\left(G^{b}\left(\tau_{j}\right)-\bar{G}\left(\tau_{j}\right)\right)
			\end{equation}
			and $\bar{G}\left(\tau_{i}\right)$ is the Monte Calro average of Green's functions of $N_b$ bins.
			
			Then we perform the Monte Carlo sampling ~\cite{Sandvik2016,Olav2008} again to optimize the spectral function. We assume that the spectral function has the following form: $A(\omega)=\sum_{i=1}^{N_{\omega}} A_{i} \delta\left(\omega-\omega_{i}\right)$ and the weight of such Monte Carlo configuration is: $
			W \sim \exp \left(-\frac{\chi^{2}}{2 \,\Theta_T}\right)$. Here $\Theta_T$ is an analogy to temperature. Then we compute the average $\langle\chi^{2}\rangle$ at different $\Theta_T$, via the simulated annealing process, at the end of it, we can choose the converged $\Theta_T$ to satisfy:
			\begin{equation}
				\langle\chi^{2}\rangle=\chi_{\min }^{2}+a \sqrt{\chi_{\min }^{2}}.
			\end{equation}
			Usually we set $a=2$, and the ensemble average of the spectra at such optimized $\Theta$ is the final one to present in the main text.
			
			We note that the QMC-SAC scheme for obtaining dynamical spectral function, is developed over the past decades and has been verified in many works on quantum many-body systems and have been directly compared with the Bethe ansatz, exact diagonalization, field theoretical analysis and spectroscopy experiments, such as the works on 1D Heisenberg chain~\cite{Sandvik2016}, 2D Heisenberg model compared with neutron scattering and field theoretical analysis~\cite{HShao2017,zhou2020amplitude}, $Z_2$ quantum spin liquid model with fractionalized spectra~\cite{GYSun2018,ZYan2021}, quantum Ising model with direct comparison with neutron scattering and NMR experiments~\cite{hu2020evidence,li2020kosterlitz}, the non-Fermi-liquid and metallic quantum critical point~\cite{jiang2020,ChuangChen2021} and the TBG system at flat-band limit~\cite{XuZhang2021}.
			
			\subsection{Section IV: Analytic Charge $\pm 1$ Excitations and Goldstone Modes (without kinetic energy)}
			
			Here we follow the Ref.~\cite{bernevig2020tbg5}. For $\nu=0$, we needn't flat metric condition, and ground state $|\Psi\rangle$ satisfies:
			\begin{equation}
				O_{\mathbf{q}+\mathbf{G}} |\Psi\rangle=0
			\end{equation}
			then:
			\begin{equation}
				\left[H_{int}, d_{\bk, n, \eta,ss}^{\dagger}\right]|\Psi\rangle=\frac{1}{2 \Omega_{\mathrm{tot}}} \sum_{m_2} R_{m_2 n}^{\eta}(\bk) d_{\bk, m_{2}, \eta,ss}^{\dagger}|\Psi\rangle
			\end{equation}
			where
			\begin{equation}
				R_{m_1 n_1}^{\eta}(\bk)=\sum_{ m,\bq,\bG,|\bq+\bG|\neq 0}V(\mathbf{q}+\mathbf{G})\lambda^{*}_{m_1,m,\eta}(\bk,\bk+\bq+\bG)\lambda_{n_1,m,\eta}(\bk,\bk+\bq+\bG) \
			\end{equation}
			
			Diagonalize $\frac{R_{m_1 n_1}^{\eta}(\bk)}{2 \Omega}$ and we obtain the charge $\pm 1$ excitations, as plotted as the dashed lines in the Fig. 2 (c) and (d) of the main text with our model parameters.
			
			The Goldstone modes can be obtained by calculating the following commutator:
			\begin{equation}
				\left[H_{int}, d_{\mathbf{k}_{2}, n_{2}, \eta_{2}, s_{2}}^{\dagger} d_{\mathbf{k}_{1}, n_{1}, \eta_{1}, s_{1}}\right]|\Psi\rangle=
				\frac{1}{2 \Omega_{\text {tot }}} \sum_{m_{2}, m_{1}} \sum_{\mathbf{q}} S_{m_{2} m_{1} ; n_{2} n_{1}}^{\left(\eta_{2}, \eta_{1}\right)}(\mathbf{k}+\mathbf{q}, \mathbf{k} ; \mathbf{p}) d_{\mathbf{k}+\mathbf{p}+\mathbf{q}, m_{2}, \eta_{2}, s_{2}}^{\dagger} d_{\mathbf{k}+\mathbf{q}, m_{1}, \eta_{1}, s_{1}}|\Psi\rangle
			\end{equation}
			
			where
			\begin{equation}
				\begin{aligned}
					S_{m_{2}, m_{1} ; n_{2}, n_{1}}^{\left(\eta_{2}, \eta_{1}\right)}(\mathbf{k}+\mathbf{q}, \mathbf{k} ; \mathbf{p})=& \delta_{\mathbf{q}, \mathbf{0}}\left(\delta_{m_{2}, n_{2}} R_{m_{1} n_{1}}^{ \eta_{1}}(\mathbf{k})+\delta_{m_{1}, n_{1}} R_{m_{2} n_{2}}^{*\;\;\eta_{2}}(\mathbf{k}+\mathbf{p})\right) \\
					&-2 \sum_{\mathbf{G}} V(\mathbf{G}+\mathbf{q}) \lambda_{n_{1},m_1,\eta_{1}}(\bk,\bk+\bq+\bG)\lambda_{n_{2},m_{2},\eta_{2}}^{*}(\mathbf{k}+\mathbf{p},\mathbf{k}+\mathbf{p}+\bq+\bG)      
				\end{aligned}
			\end{equation}
			Here, we treat $(\bk+\bq, m_1, m_2)$ as one subindex, and we will call it $i$. The second index $(\bk,n_1,n_2)$ will be called $j$. So we have a matrix $S^{\left(\eta_{2}, \eta_{1}\right)}_{ij} (\bkp)$ for each $\bkp$, where $S$ is a $4N*4N$ matrix. Diagonalize $\frac{S^{\left(\eta_{2}, \eta_{1}\right)}_{ij} (\bkp)}{2 \Omega}$ and we obtain the Goldstone modes, as plotted as the dashed lines in the Fig. 2 (e) and (f) of the main text with our model parameters.
	\end{widetext}
\end{appendix}
		
\end{document}